\documentclass[11pt,letterpaper]{article}
\pdfoutput=1
\usepackage{jheppub}
\usepackage{bbm}
\usepackage{mathrsfs}
\usepackage{slashed}
\usepackage{caption}
\usepackage{epstopdf}
\usepackage[normalem]{ulem}
\usepackage[bottom]{footmisc}
\usepackage{subcaption}
\usepackage{bbold}
\usepackage{titlesec}
\usepackage{threeparttable}
\usepackage{booktabs}
\usepackage{changepage}
\usepackage[utf8]{inputenc}
\usepackage{dsfont} 

\usepackage{grffile}
 
\usepackage{graphicx}  % needed for figures
\usepackage{dcolumn}   % needed for some tables
\usepackage{bm}        % for math
\usepackage{amssymb}   % for math
\usepackage{setspace}
\usepackage{amsmath, amssymb, setspace}
\usepackage{array}
\usepackage{booktabs}
\usepackage{caption}
\usepackage{indentfirst}
\usepackage{float}
\usepackage{lmodern}
\usepackage{multirow}
\usepackage{soul}
\usepackage[normalem]{ulem}

\usepackage{braket}
\usepackage{comment}

\usepackage[draft]{pgf}

\usepackage{adjustbox} 
\newlength{\myimageoversize}
\newsavebox{\myimage}

\usepackage[most]{tcolorbox}

\usepackage{empheq}

\tcbset{colback=white!10!white, colframe=black!50!black, 
        highlight math style= {enhanced, %<-- needed for the ’remember’ options
            colframe=black,colback=white!10!white,boxsep=0pt}
        }

\titleformat{\subsection}
 {\normalfont\fontsize{12}{17}\itshape}{\thesubsubsection}{1em}{}
 
\title{\huge{Cosmological Dependence of Non-resonantly Produced Sterile Neutrinos}}

\author[a]{Graciela B. Gelmini,}
\author[a]{Philip Lu}
\author[a]{and Volodymyr Takhistov}

\affiliation[a]{Department of Physics and Astronomy, UCLA,\\
475 Portola Plaza, Los Angeles, CA 90095, USA}

\emailAdd{gelmini@physics.ucla.edu}
\emailAdd{philiplu11@gmail.com}
\emailAdd{vtakhist@physics.ucla.edu}

\abstract{We discuss how a laboratory detection of a  sterile neutrino not only would constitute a fundamental discovery of a new particle, but could also provide an indication of the evolution of the Universe before Big-Bang Nucleosynthesis (BBN), a fundamental discovery in cosmology. These ``visible" sterile neutrinos could be detected in experiments such as KATRIN/TRISTAN and HUNTER in the keV mass range and PTOLEMY, KATRIN and reactor neutrino experiments in the eV mass range.  Standard assumptions are usually made to compute the relic abundance and momentum distribution of particles produced before the temperature of the Universe was 5 MeV, an epoch from which there are no observed remnants thus far. However, non-standard pre-BBN cosmologies based on other assumptions that are equally in agreement with all existing data can arise in some theoretical models. We revisit the  production of 0.01 eV to 1 MeV sterile neutrinos via non-resonant active-sterile flavor oscillations in several pre-BBN cosmologies.  We give general equations for models in which the expansion of the Universe is parametrized by its amplitude and temperature power and where  entropy is conserved, which include kination and scalar tensor models as special cases.
}

\begin{document}
\preprint{}
 \maketitle
\flushbottom

\section{Introduction}

The earliest detected cosmological remnants 
are the light nuclei produced during Big Bang Nucleosynthesis (BBN). The lower limit on the highest temperature of the radiation-dominated epoch in which BBN happened is just close to 5 MeV~\cite{deSalas:2015glj, Hasegawa:2019jsa, DeBernardis:2008zz, Hannestad:2004px, Kawasaki:2000en, Kawasaki:1999na}. Thus, the cosmological evolution in the Universe before the temperature of the Universe was 5 MeV is unknown. 

Many dark matter (DM) particle candidates are produced before the temperature of the Universe was $T =$ 5 MeV, thus assumptions must be made about cosmology in that epoch to compute their relic abundance and momentum distribution.  The standard assumptions that are  usually made, which constitute the ``standard pre-BBN cosmology", are that the Universe was radiation-dominated and only Standard Model (SM) particles are present. Additionally, it is assumed that no extra entropy in matter and radiation is produced. This is an extension of the known cosmology    below  5 MeV to higher temperatures. However, there are cosmological models with alternative assumptions that are equally in agreement with all existing data. This is the case in some well motivated theoretical models, based e.g. on moduli decay,  quintessence and extra dimensions.~A non-standard cosmological evolution could drastically alter the relic density an momentum distribution of any relics produced before the temperature of the Universe was 5 MeV. Detecting any of these relics would open a new window into
  this yet unexplored cosmological epoch.  

There are three active neutrinos $\nu_\alpha$, characterized by their flavors $\alpha = e, \mu, \tau$, coupled to the W and Z weak gauge bosons\footnote{Measurements of the invisible decay width of the Z boson limit the number of weakly interacting neutrino species with mass below 45 GeV to three~\cite{ALEPH:2005ab}.}. These neutrinos are massless within the SM,  however, neutrino oscillations~\cite{Fukuda:1998mi} show that they are massive. This motivates the study  of scenarios with more than three neutrino flavors, as many  neutrino mass models include  one or more additional ``sterile'' neutrinos $\nu_s$ that do not directly couple to the W and Z bosons. 
 
The theory of active neutrino oscillations has been extensively tested by experimental measurements of neutrino production in the Sun~\cite{Cleveland:1998nv,Abdurashitov:2009tn,Altmann:2005ix,Hampel:1998xg,Aharmim:2005gt,Abe:2016nxk}, in nuclear reactors~\cite{Araki:2004mb,An:2012eh,Abe:2013sxa,Ahn:2012nd}, at accelerators~\cite{Ahn:2006zza,Adamson:2013whj,Abe:2014ugx,Agafonova:2014bcr,Adamson:2016xxw} and in the atmosphere~\cite{Aartsen:2014yll,Fukuda:1998mi}.
While the data has been generally consistent with the three-flavor paradigm, several experiments reported anomalies that could be consistently explained by introducing one or more additional sterile neutrinos into the theory with a mass of $m_{s}=\mathcal{O}$(eV).
In particular, the short-baseline experiments\footnote{In short-baseline experiment the detectors are located at less than 1 km away from the source.} LSND~\cite{Aguilar:2001ty} and MiniBooNE~\cite{Aguilar-Arevalo:2013pmq} reported excesses of $\bar{\nu}_e$ within $\bar{\nu}_\mu$ beams, and MiniBooNE also found an excess in $\nu_e$ appearance~\cite{Aguilar-Arevalo:2018gpe}.  A deficit of $\nu_e$ flux from radioactive calibration sources has been observed in gallium experiments\footnote{Recent re-evaluations of gallium cross-sections indicate a weaker sterile neutrino preference~\cite{Kostensalo:2019vmv}.}~\cite{Bahcall:1994bq,Abdurashitov:2005tb}. Furthermore, an under-abundance of $\bar{\nu}_e$ has been reported in reactor neutrino experiments\footnote{This anomaly has weakened in light of Daya Bay’s reactor fuel cycle measurements~\cite{An:2017osx} and observations of spectral distortions not predicted by flux calculations~\cite{Huber:2016xis}.}~\cite{Mention:2011rk}. Additionally, combined fits~\cite{Dentler:2018sju,Gariazzo:2018mwd,Liao:2018mbg} to recent reactor neutrino results by the DANSS~\cite{Alekseev:2018efk} and NEOS~\cite{Ko:2016owz} experiments are consistent with an interpretation based on  a sterile neutrino with a mass of
$m_{s}=\mathcal{O}$(eV) and mixing with a $\nu_e$ active neutrino. 
Reactor neutrino data from the Daya Bay~\cite{An:2016luf}, Bugey-3~\cite{Declais:1994su} and  PROSPECT~\cite{Ashenfelter:2018iov} experiments  constrain these sterile neutrinos and the PTOLEMY~\cite{Betti:2019ouf} and KATRIN~\cite{megas:thesis}  experiments will be able to test all or part of this parameter space.

Sterile neutrinos with mass of $\mathcal{O}$(keV) and a spectrum close to thermal constitute a viable Warm DM (WDM) candidate (see e.g. Ref.~\cite{Boyarsky:2018tvu}).  The mass and mixing of sterile neutrinos that make up all or most of the DM is subject to astrophysical constraints, such as those imposed by Lyman-$\alpha$ forest and X-ray observations. If the keV-mass sterile neutrinos are produced from sterile-active oscillations in the early Universe, these limits disfavor them as the sole DM component (see e.g Ref~\cite{Boyarsky:2008ju}). These bounds are drastically weakened if sterile neutrinos constitute a sub-dominant DM component (see e.g. Ref.~\cite{Palazzo:2007gz}).  It has been suggested that the $3.5$ keV X-ray emission line observed in 2014~\cite{Bulbul:2014sua,Boyarsky:2014jta}  could be produced in the decay of $m_s = 7$ keV sterile neutrinos. The KATRIN laboratory experiment with its proposed TRISTAN upgrade~\cite{Mertens:2018vuu}, as well as the upcoming HUNTER experiment and its upgrades~\cite{Smith:2016vku}, will test sterile neutrinos in the keV-scale mass range.

If produced in a  supernova explosion, sterile neutrinos with mass larger than keV could carry away a sizable fraction of the emitted energy.  Asymmetric emission of the sterile neutrinos due to the presence of a strong magnetic field could explain the observed large velocities of pulsars~\cite{Fuller:2003gy}. These effects are independent of the relic abundance of sterile neutrinos and the fraction of the DM they constitute. 

Sterile neutrinos without additional interactions beyond the SM  that couple to the SM particles only through mixing with active neutrinos,  as we assume here, are produced in the early Universe through active-sterile  flavor oscillations and collisional processes.
 For simplicity we assume a $\nu_s$ that mixes with only one of the $\nu_{\alpha}$, which for our figures is $\nu_e$, with a mixing of $\sin \theta$. In the absence of a large lepton asymmetry the oscillations are non-resonant, and the resulting relic number density was first obtained by Dodelson and Widrow~\cite{Dodelson:1993je}. In the standard cosmology this mechanism results in a Fermi-Dirac momentum distribution of sterile neutrinos, with a reduced magnitude with respect to active neutrinos. In the presence of a significant lepton asymmetry in active neutrinos, sterile neutrinos are  instead produced via resonant oscillations, as pointed out by Shi and Fuller~\cite{Shi:1998km}. Then, the resulting sterile neutrinos have a colder momentum distribution  (i.e. with a lower average momentum) that is different from a Fermi-Dirac spectrum. This mechanism was studied in its generality within Ref.~\cite{Abazajian:2001nj,Abazajian:2004aj}. 
In non-minimal particle models, sterile neutrino production could also proceed via other mechanisms, such as decays of additional heavy scalars~\cite{Petraki:2007gq}.

In this work we revisit the effects of different pre-BBN cosmologies on sterile neutrinos with mass $10^{-2}$ eV $< m_s <$ 1 MeV, produced via non-resonant  active-sterile oscillations. Several related studies have
been previously carried out~\cite{Gelmini:2004ah,Rehagen:2014vna,Abazajian:2017tcc}.  We update the older constraints of Ref.~\cite{Gelmini:2004ah}, extend the results of Ref.~\cite{Rehagen:2014vna} -- pointing out  the significance of upcoming laboratory experiments, and extend the analysis of Ref.~\cite{Abazajian:2017tcc} on keV-mass neutrinos  down to 0.01 eV masses. Here, we provide additional details and further expand on the results presented in our recent letter~\cite{Gelmini:2019esj}.

In cosmological models in which entropy is conserved the sterile neutrino production depends crucially 
on the magnitude and temperature dependence of the Hubble expansion rate $H$. If in the non-standard cosmological phase $H$ is larger than in the standard cosmology, the production of sterile neutrinos during this phase is suppressed. Notably, a particular scalar-tensor model , which was not discussed in the previous study of Ref.~\cite{Rehagen:2014vna}, allows for a lower expansion rate compared to the standard cosmology and the production is enhanced.  We assess the effects of these considerations on the  possibility of detecting a sterile neutrino in laboratory experiments.
For comparison, we also  reconsider low reheating temperature models~\cite{Gelmini:2004ah}, in which 
 entropy is produced and the radiation bath is subdominant during the non-standard phase. Hence, the dominant sterile neutrino production in this scenario occurs during the late standard cosmological phase. 
 
This paper is organized as follows. In Section~\ref{sec:modcos} we describe a general parametrization for the expansion rate of the Universe $H$ for models in which we assume that the entropy in matter and radiation is conserved. In Section~\ref{sec:nonres} we present the effects of different  pre-BBN cosmologies on non-resonant sterile neutrino production.  In Section~\ref{sec:limits} we describe the resulting sterile neutrino limits and regions of interest in mass-mixing space. Finally, in Section~\ref{sec:summary} we summarize our results.
 
\section{Early Universe cosmology}
\label{sec:modcos}

The  expansion rate of the Universe, the Hubble parameter $H = (\dot{a}/a)$ -- where $a$ is the cosmological scale factor of the Universe, is determined by the Friedmann equation. In the  standard  cosmological model~\cite{Kolb:1990vq} the Universe was radiation dominated before BBN, and the highest  temperature $T$ of the radiation bath achieved in this epoch is  much higher than the temperature $T \simeq 0.8$ MeV at which BBN starts. In the standard cosmology $H$ is
\begin{equation} 
\label{eq:hStd}
H_{\rm Std} = \sqrt{\dfrac{8 \pi G \rho(T)}{3} } =  \Big(\dfrac{T^2}{M_{\rm Pl}}\Big)  \sqrt{\dfrac{8 \pi^3 g_{\ast}(T)}{90}}~.
\end{equation}
Here $\rho(T)= (\pi^2/30) g_{\ast}(T) T^4$ is the total energy density, $M_{\rm Pl}= 1.22 \times 10^{19}$ GeV is the Planck mass and $g_\ast(T)$ is the number of degrees of freedom contributing to the energy density at temperature $T$. Assuming that only SM particles are present for $T$ higher than the QCD phase transition at $T \simeq 200$ MeV, we have  $g_\ast=80$ and  it is approximately constant (it reaches $g_\ast=$100 at $T \simeq $100 GeV). Close to the QCD phase transition, the value of $g_\ast$ decreases steeply with decrement of $T$, and we take a characteristic value of $g_{\ast} \simeq 30$ until $T$ decreases to $T= 20$ MeV. Between this temperature and $T =1$ MeV, when electrons and positrons become non-relativistic and annihilate, $g_{\ast} = 10.75$ (see e.g. Refs. \cite{Husdal:2016haj,Borsanyi:2016ksw,Drees:2015exa}). Unless  otherwise stated,  for simplicity we use $g_\ast=30$ in our figures.

\subsection{Non-standard pre-BBN cosmologies}

The requirement of a successful BBN, which also insures that the subsequent history of the Universe develops as usual, imposes that the Universe is radiation dominated for temperatures $T \lesssim 5$ MeV~\cite{deSalas:2015glj, Hasegawa:2019jsa, DeBernardis:2008zz, Hannestad:2004px, Kawasaki:2000en, Kawasaki:1999na}. However, a non-standard cosmological evolution is allowed at higher temperatures. 

Any new additional contribution to the energy density in matter or radiation, or equivalently to the geometry sector of Einstein's equations, results in modification of the Hubble expansion rate through the Friedmann equation. Except for the low reheating temperature model (see below),  we consider non-standard cosmologies in which the entropy in matter and radiation is conserved (and hence, the relation between the scale factor of the Universe $a$ and the temperature  $T$ follows $a \sim T^{-1}$), but the Hubble expansion $H$ as a function of $T$ is non-standard. In all cosmologies of this type that we will consider $H$ can be given by a simple  parameterization~\cite{Catena:2009tm} 
\begin{equation} \label{eq:hnStd}
    H = \eta~ \Big(\dfrac{T}{T_{\rm tr}}\Big)^{\beta} H_{\rm Std}~,
\end{equation}
where $T_{\rm tr}$ is a reference temperature, which we identify with the temperature at which before BBN the cosmology transitions to the standard  cosmology,  $\eta$ and  $\beta$ are real parameters and $\eta$ is positive.

To preserve BBN, we require that $H_{\rm Std}$ is recovered in Eq.~\eqref{eq:hnStd} at $T< T_{\rm tr}=$ 5 MeV. Various phenomenological studies of non-standard cosmologies have been 
carried out~\cite{Gelmini:2010zh,Catena:2009tm,Kamionkowski:1990ni,Salati:2002md,Lambiase:2018yql}, which can be generally classified by the value of the $\beta$ parameter: $\beta > 2$ for ultra-fast expansion, e.g. when the Universe is dominated by a field with an exponential potential~\cite{DEramo:2017gpl} (as in the ekpyrotic scenario~\cite{Khoury:2001wf}), $\beta = 2$ in the Randall-Sundrum type II brane cosmology~\cite{Randall:1999vf} (see discussion in Ref.~\cite{Schelke:2006eg}), $\beta =$ 1 or larger was considered in ``fast-expanding" models in which there is  an additional energy density component from a non-interacting component~\cite{DEramo:2017gpl}, $\beta = 1$ in kination models~\cite{Spokoiny:1993kt,Joyce:1996cp,Salati:2002md,Profumo:2003hq,Pallis:2005hm,Schelke:2006eg},  $\beta = 0$ in cosmologies with an overall boost of the Hubble expansion rate, e.g. with a large number of additional relativistic degrees of freedom within the thermal plasma~\cite{Catena:2009tm}, $\beta = -0.8$ can occur in variants of the scalar-tensor cosmology~\cite{Catena:2009tm,Catena:2004ba} and $\beta = 2/n - 2$ describes $f(R)$ gravity\footnote{For $n = 2$ this reduces to the Starobinsky model~\cite{Starobinsky:1980te}.}, with $f(R) = R + const. \times R^n$~\cite{Capozziello:2015ama}.

We will provide expressions in terms of $\eta$ and $\beta$ for all the relevant equations in our study, but we will focus our discussion on two often considered modified cosmologies, kination (K) and scalar-tensor (ST) models.

\subsection{Kination (K)}

In the kination phase~\cite{Spokoiny:1993kt,Joyce:1996cp,Salati:2002md,Profumo:2003hq,Pallis:2005hm}, the kinetic energy of a scalar field  $\phi$ dominates over its potential energy and all other contributions to the total energy density $\rho_{tot}$. Hence, it also governs the expansion rate in the early Universe. Cosmologies with phases governed by such ``fast-rolling'' scalar fields can arise in models of quintessence.

During the kination period  $\rho_{tot}\simeq \rho_\phi \simeq \dot{\phi}^2/2 \sim a^{-6}$, where $\rho_{\phi}$ is the energy density of the scalar $\phi$, the associated expansion rate of the Universe during the kination phase is $H_{\rm K}\sim \sqrt{\rho_{tot}}\sim T^3$.  The ratio of $\phi$-to-photon energy density at $T \simeq 1$ MeV, $\eta_\phi=\rho_\phi/\rho_\gamma$, fixes the contribution of the $\phi$ kinetic energy to the total energy density at higher temperatures. Hence, one has $H_{\rm K}\simeq \sqrt{\eta_\phi}(T/1\,{\rm MeV})H_{\rm Std}$.  The value of $\eta_\phi$ can be determined by assuming a rapid transition from the kination phase to the radiation dominated phase at the transition temperature $T_{tr}$, so that $H_{\rm K}(T_{tr})=H_{\rm Std}(T_{tr})$. With this approximation  the expansion rate of the Universe during the kination phase is 
\begin{equation} 
\label{eq:hkin}
H_{\rm K}= \Big(\dfrac{T^3}{M_{\rm Pl} T_{\rm tr}}\Big) \sqrt{\dfrac{8\pi^3 g_{\ast}}{90}}
= \Big(\dfrac{T}{T_{\rm tr}}\Big) ~ H_{\rm Std}~~,
\end{equation}
which in Eq.~\eqref{eq:hnStd} corresponds to $\eta = 1$ and $\beta = 1$.

\subsection{Scalar-tensor (ST1 and ST2)}

Scalar-tensor models of gravity \cite{Santiago:1998ae,Catena:2004ba} have one or more scalar fields coupled through the metric tensor to the matter sector. These extra fields affect the expansion rate of the Universe when the temperature of the thermal bath is higher than the transition temperature $T_{tr}$, at which a fast transition is assumed to occur before BBN, so that the theory becomes indistinguishable from General Relativity  at $T < T_{tr}$. Such scalar-tensor modified cosmologies can appear in models of extra dimensions (e.g.~~\cite{Koivisto:2013fta}). Depending on the details of the scenario, the respective early Universe expansion rate $H_{\rm ST}$ can be either larger~\cite{Catena:2004ba} or slightly smaller ~\cite{Catena:2007ix}  than the expansion rate within the standard cosmology $H_{\rm Std}$. 

As benchmarks, we consider two scalar-tensor models that are extreme in terms of the magnitude of the expansion rate they predict, which we call ST1 and ST2 and assume $T_{\rm tr} = 5$ MeV. In the ST1 model from Ref.~\cite{Catena:2004ba} the expansion rate is enhanced compared to the standard cosmology, described by
\begin{equation} \label{eq:hst1}
H_{\rm ST1} = 7.4 \times 10^5 \Big(\dfrac{ T_{\rm tr}^{0.8}~T^{1.2}}{M_{\rm Pl}}\Big)   \sqrt{\dfrac{8\pi^3 g_{\ast}}{90}} = 7.4 \times 10^5 \Big(\dfrac{T}{T_{\rm tr}}\Big)^{-0.8} ~ H_{\rm Std}~,
\end{equation}
which in Eq.~\eqref{eq:hnStd} corresponds to $\eta = 7.4 \times 10^5$ and $\beta = -0.8$.

In Ref.~\cite{Catena:2007ix} it was shown that
contributions of an additional ``hidden'' matter sector beyond the visible sector can result in a reduced expansion rate, compared to the standard cosmology. We choose for ST2 a model of this type with the lowest expansion rate found in Ref.~\cite{Catena:2007ix}, $(H_{\rm ST}/H_{\rm Std})^2= 10^{-3}$, and for which we choose $\beta=0$ (given the  variations in the behavior of H close to $T_{\rm tr}$ of the numerical solutions shown in Fig. 4 of Ref.~\cite{Catena:2007ix}). Thus
\begin{equation} 
\label{eq:hst2}
H_{\rm ST2} = 3.2 \times 10^{-2} \Big(\dfrac{1}{M_{\rm Pl}}\Big) T^{2}  \sqrt{\dfrac{8\pi^3 g_{\ast}}{90}}= 0.03 ~ H_{\rm Std}~,
\end{equation}
which in Eq.~\eqref{eq:hnStd} corresponds to $\eta = 0.03$ and $\beta = 0$.

\begin{figure}
\begin{center}
\includegraphics[scale=.75]{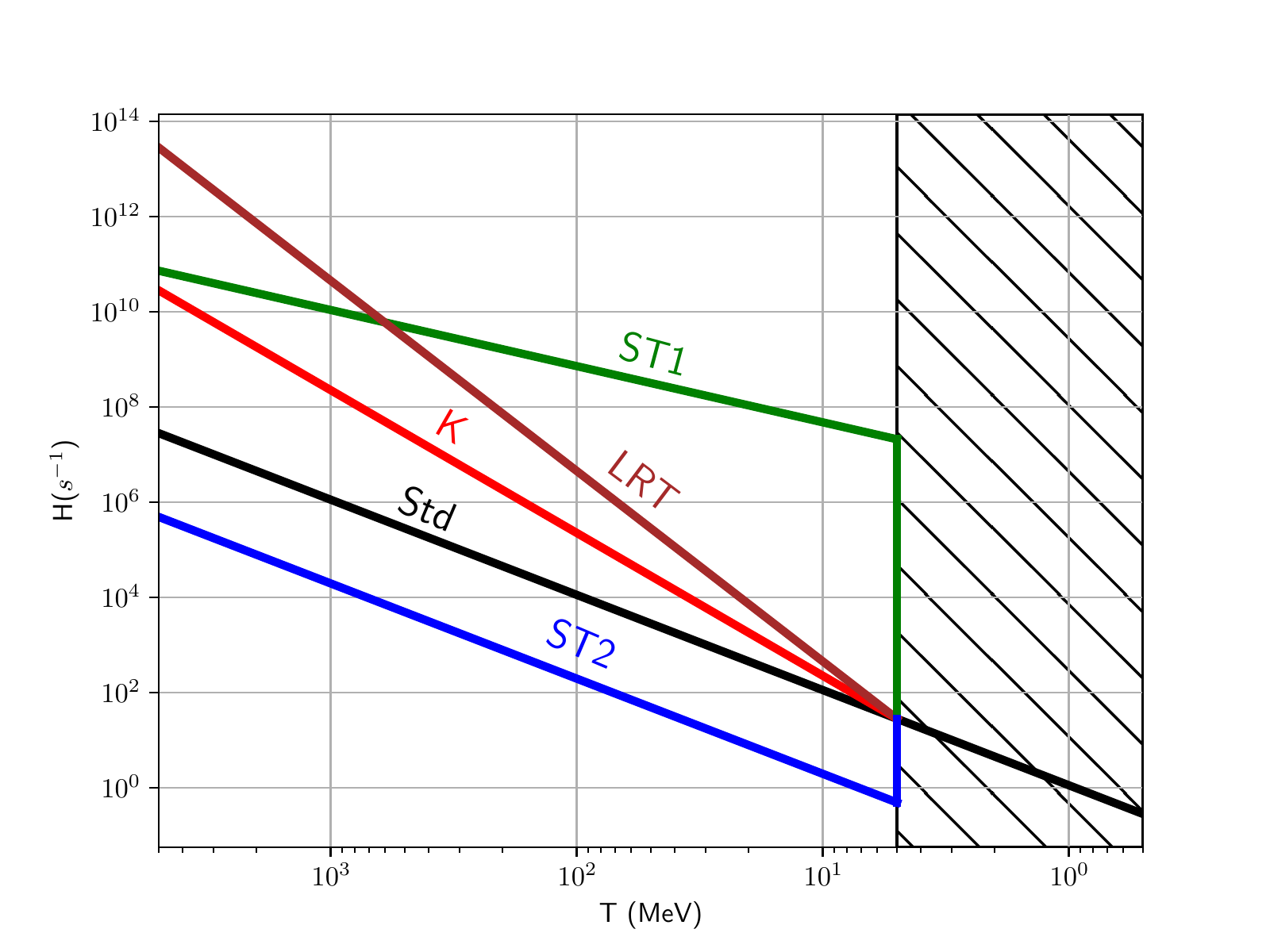}
\caption{Expansion rate of the Universe $H$ as a function of  the temperature $T$ of the radiation bath for the Std (black), K (red), ST1 (green) and ST2 (blue) and LRT (brown) cosmologies.  At $T_{tr}=5$ MeV, the upper boundary of the hatched region, all the non-standard cosmologies transition to the standard cosmology. For simplicity, we assume the transition to be sharp in  the ST1 and ST2, cosmologies.   }
\label{fig:hubble}
\end{center}
\end{figure}

 Expansion rates in between $H_{\rm ST1}$ and $H_{\rm ST2}$ can also appear within scalar-tensor models and our analysis can be readily applied to them.

 Fig.~\ref{fig:hubble} shows the expansion rate of the Universe $H$ as a function of the temperature $T$ of the radiation bath  for the Std,  $H_{\rm Std}$ (black), K, $H_{\rm K}$ (red), ST1, $H_{\rm ST1}$ (green), and ST2, $H_{\rm ST2}$ (blue), cosmologies assuming a fast transition at $T_{tr}=5$ MeV and that for $T< T_{tr}$ the  cosmology is standard. 
 
 We note that since we impose that non-standard cosmologies transition to the standard cosmology before the onset of BBN, restrictions arising from consistency with current astrophysical observations (e.g. signals from neutron star binary mergers~\cite{Sakstein:2017xjx}) do not affect our considerations.

\subsection{Low reheating temperature (LRT)}

The phenomenological parametrization of Eq.~\eqref{eq:hnStd} does not capture the whole modification to cosmology in models in which the entropy in radiation and matter is not conserved and consequently the temperature $T$ dependence on the scale factor $a$ is different than the usual $T \sim 1/a$. One of these is the low reheating temperature (LRT) model. 

In the LRT model a scalar field $\phi$ oscillates coherently around its true minimum and dominates the energy density of the Universe. Decays of $\phi$ produce a radiation bath that thermalizes  to a temperature $T$ and becomes dominant at the reheating temperature $T_{\rm RH}$. Subsequently, the radiation dominates the energy density of the Universe  for $T < T_{\rm RH}$ (see e.g. Refs.~\cite{Gelmini:2006pw, Gelmini:2006pq}).  Such field $\phi$ could be the inflaton itself or a modulus field producing a late episode of entropy production~(see e.g.~Ref.~\cite{Moroi:1994rs, Kawasaki:1995cy, Moroi:1999zb,Chen:2018uzu, Kitano:2008tk,Drees:2017iod}).   Other alternative possibilities include Q-ball decays~(e.g.~\cite{Fujii:2002kr}). 

In Fig.~\ref{fig:hubble}, we show the expansion rate for LRT  with $T_{\rm RH}= T_{\rm tr} = 5$ MeV in brown.

The LRT scenario  may drastically alter the DM relic abundance (see e.g. \cite{Gelmini:2004ah, Gelmini:2006pq, Gelmini:2006pw, Yaguna:2007wi, Gelmini:2008fq} and discussion in Ref.~\cite{Gelmini:2010zh}). The non-resonant production of sterile neutrinos in LRT models was considered in  Refs.~\cite{Gelmini:2004ah, Gelmini:2008fq}, with the assumption that it predominantly occurs during the standard cosmological phase, when $T < T_{\rm RH}$, since
the thermal bath is subdominant before reheating.  This approximation was validated by considering also the production during the non-standard phase in Ref.~\cite{Yaguna:2007wi}. Because the non-resonant production rate is far from its maximum for $T < T_{\rm RH}$, the relic abundance of sterile neutrinos is suppressed in these models. In this study we reconsider the production of sterile neutrinos in  LRT model  with $T_{\rm RH} =T_{\rm tr}= 5$ MeV~\cite{Gelmini:2004ah, Yaguna:2007wi}  and update the observational and experimental bounds on them.

\section{Non-resonant sterile neutrino production}
\label{sec:nonres}

In the absence of a significant primordial lepton asymmetry, the production of sterile neutrinos happens via non-resonant flavor oscillations between the active neutrinos $\nu_\alpha$ of the SM and the sterile neutrino $\nu_s$. This is  called the Dodelson-Widrow mechanism (DW)~\cite{Dodelson:1993je}  because  they derived the analytic solution for the relic number density of sterile neutrinos produced in this manner\footnote{The original Dodelson-Widrow results were subsequently corrected by a factor of 2~\cite{Dolgov:2002wy, Abazajian:2001nj}.}  (see also earlier work~\cite{Barbieri:1989ti,Kainulainen:1990ds}).
Interactions of active neutrinos with the surrounding plasma during  the oscillations act as measurements and cause the collapse of the wave function into one of the oscillating states, which with some probability results in a sterile neutrino. The production rate is usually not fast enough for sterile neutrinos to equilibrate and the process is a freeze-in of the final abundance.

\subsection{Boltzmann equation}
\label{ssec:boltzmann}

  Assuming that only two neutrinos mix, $\nu_s$ and one active neutrino $\nu_\alpha$  (which we assume to be  $\nu_e$ in our figures), the time evolution of the phase-space density distribution function of sterile neutrinos $f_{\nu_s}(p,t)$ with respect to the density function of active neutrinos $f_{\nu_{\alpha}}(p,t)$ is given by the following Boltzmann equation~\cite{Kolb:1990vq,Abazajian:2001nj}
\begin{align} 
\label{eq:boltzmann}
    \frac{d}{dt}f_{\nu_s}(p,t) ~=&~ \frac{\partial}{\partial t}f_{\nu_s}(p,t) - Hp\frac{\partial}{\partial p}f_{\nu_s}(p,t) \notag\\ ~=&~ \Gamma(p,t) \Big[f_{\nu_\alpha}(1-f_{\nu_s})-f_{\nu_s}(1-f_{\nu_\alpha})\Big]~.
\end{align}
Here $H$ is the expansion rate of the Universe, $p$ is the magnitude of the neutrino momentum and $\Gamma (p,t)$ is the conversion rate of active to sterile neutrinos.
The active neutrinos are assumed to have a Fermi-Dirac distribution
\begin{equation} 
\label{eq:fermid}
 f_{\nu_\alpha} = (e^{\epsilon-\xi}+1)^{-1}~,   
\end{equation}
where $E=p$ because all neutrinos we are interested in are relativistic during the production, $\epsilon = p/T$ is the  $T$-scaled dimensionless momentum,  and $\xi = \mu_{\nu_{\alpha}}/T$ is the $T$-scaled dimensionless chemical potential with $\mu_{\nu_{\alpha}}$ being the chemical potential of  active $\nu_{\alpha}$. In the DW mechanism $\mu_{\nu_{\alpha}}$ is negligible and so we take $\xi=0$. Since $f_{\nu_s} \ll 1$, we can ignore Pauli blocking, i.e. $(1-f_{\nu_s})=1$,  and while $f_{\nu_s} \ll f_{\nu_{\alpha}}$ we can also neglect the second term on the right hand side of Eq.~\eqref{eq:boltzmann}. Changing variables, Eq.~\eqref{eq:boltzmann} can be further recast into a more convenient form~\cite{Rehagen:2014vna,Abazajian:2001nj}
\begin{equation}
\label{eq:boltzmann2}
    -HT\left(\frac{\partial f_{\nu_s}(E,T)}{\partial T}\right)_{E/T=\epsilon} \simeq~ \Gamma(E,T)f_{\nu_\alpha}(E,T)~,
\end{equation}
where  the derivative on the left-hand side is computed at constant $\epsilon$.

The conversion rate $\Gamma$ is the total interaction rate $\Gamma_{\alpha} = d_{\alpha} G_F^2 \epsilon T^3$ of the active neutrinos with the surrounding plasma weighted by the average active-sterile oscillation probability $\langle P_m \rangle$  in matter  (see Eq.~(6.5) and (6.5) of Ref.~\cite{Abazajian:2001nj})
\begin{equation} 
\label{eq:interaction}
    \Gamma ~=~ \dfrac{1}{2}  \langle P_m (\nu_\alpha \rightarrow \nu_s) \rangle \Gamma_{\alpha}   ~\simeq~ \frac{1}{4}\sin^2(2\theta_m)d_\alpha G_F^2\epsilon T^5~.
\end{equation}
In this equation $\theta_m$ is the active-sterile mixing angle in matter and $d_{\alpha}$ is a flavor-dependent parameter, 
 $d_\alpha = 1.13$ for $\nu_e$ and $d_\alpha = 0.79$ for $\nu_{\mu}$, $\nu_{\tau}$. Taking into account contributions from the thermal potential $V_T$ and the density potential $V_D$ (that is proportional to the lepton number), the matter mixing angle is given by~\cite{Abazajian:2001nj}
\begin{equation}
\label{eq:mattermixing}
    \sin^2(2\theta_m) = \frac{\sin^2(2\theta)}{\sin^2(2\theta) + \Big[\cos(2\theta)-2\epsilon T (V_D+V_T)/m_s^2\Big]^2}~.
\end{equation}
Here, the quantum damping  term in the denominator  has been omitted, because it is always negligible
for the cases we consider.

For DW production, the density potential $V_D$ is assumed to be negligible. The thermal potential $V_T$ is given by
\begin{equation}
\label{eq:thermalpotential}
    V_T = -B\epsilon T^5~,
\end{equation}
where the prefactor $B$ depends on the active neutrino flavor (indicated in parenthesis in the following equation)  and on the temperature range (indicated to the right),
\begin{align} \label{eq:bprefac}
\begin{split}
    B &= \left\{
                \begin{array}{llc}
                  10.88\times10^{-9}~\textrm{GeV}^{-4}~(e);~
                  &3.02\times10^{-9}~\textrm{GeV}^{-4}~(\mu, \tau);   &T\lesssim20\textrm{ MeV}~\\
                  10.88\times10^{-9}~\textrm{GeV}^{-4}~(e, \mu);  &3.02\times10^{-9}~\textrm{GeV}^{-4}~(\tau); ~~~~~ 20\textrm{ MeV} \lesssim &T \lesssim 180 \textrm{ MeV} \\
                  10.88 \times10^{-9}~\textrm{GeV}^{-4}~(e,\mu,\tau);  & &T\gtrsim180 \textrm{ MeV}
                \end{array}
              \right.
\end{split}
\end{align}

Since the sterile neutrino production rate $(\partial f_{\nu_s}/\partial T)_\epsilon$ is inversely proportional to the expansion rate $H$ (see Eq.~\eqref{eq:boltzmann2}),  high values of $\eta>1$ in Eq.~\eqref{eq:hnStd} result in suppressed sterile neutrino production, for fixed $m_s$ and $\sin^2(2\theta)$. Hence,  non-standard cosmological models with large  $\eta \gg 1$ are less constrained by cosmological and astrophysical upper limits on the relic density, and thus larger active-sterile mixing angles become allowed by these limits. The enlarged open parameter space for the K and ST1   (and also LRT) models places visible sterile neutrinos with $m_s \simeq $ keV
within closer reach of laboratory experiments such as KATRIN~\cite{Mertens:2018vuu} and HUNTER~\cite{Smith:2016vku}. The effect of ST1 is particularly pronounced and there are no astrophysical or cosmological limits on visible $\nu_s$ with $m_s \simeq $ eV, which are  tested in reactor and accelerator experiments.  On the other hand, the ST2 model of Eq.~\eqref{eq:hst2} with $\eta < 1$ can produce all of the DM at a smaller mixing angle for a given mass than in the standard cosmological scenario.

\subsection{Temperature of maximum non-resonant production}

\begin{figure}
\begin{center}
\includegraphics[scale=.45]{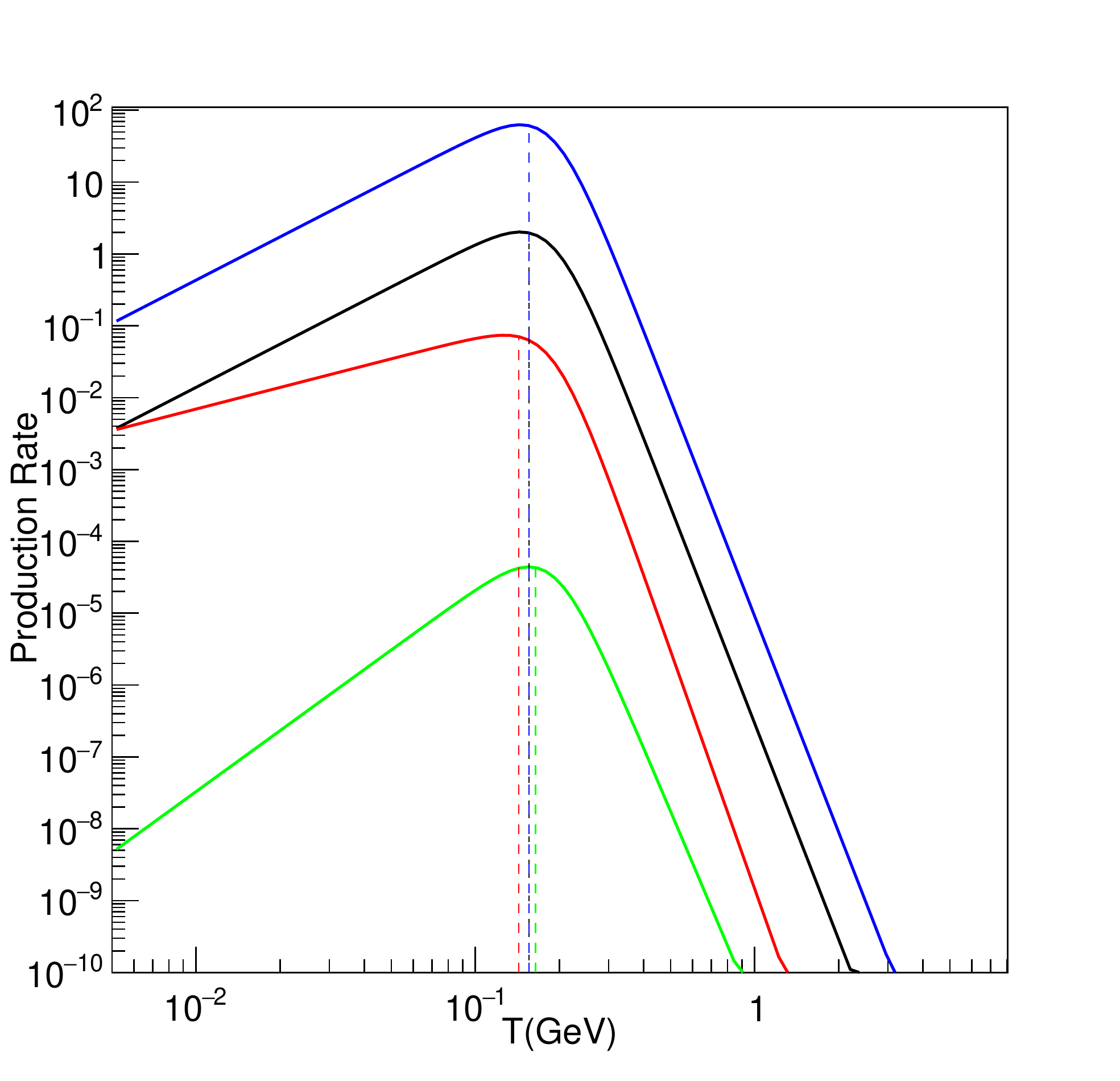}
\caption{Sterile neutrino non-resonant production rate $({\partial f_{\nu_s}(E,T)}/{\partial T})_{\epsilon}$ in Eq.~\eqref{eq:boltzmann2} as  function of the temperature $T$ for $\epsilon = 1$ and $m_s = 1$ keV in the Std (black), K (red), ST1 (green) and ST2 (blue) cosmologies, clearly showing their inverse proportionality with the magnitude of the expansion rate $H$ and also minor differences in shape and width due to the different values of the $\beta$ parameter. The value of $T_{\rm max}$ in each case is indicated by a vertical dashed line of the color of the corresponding cosmology. }
\label{fig:transitions}
\end{center}
\end{figure}

The sterile neutrino conversion rate $\Gamma$, given in Eq.~\eqref{eq:interaction} and Eq.~\eqref{eq:mattermixing}, is suppressed at high temperatures by the thermal potential $V_T$, so that $\Gamma \sim T^{-7}$. At low temperatures, where $V_T$ is negligible, it is suppressed by the decreasing interaction rate, so that $\Gamma \sim T^5$. Where both regimes cross, for cosmologies with $\beta < 2$, the production rate $(\partial f_s/\partial T)_{\epsilon}$ has a narrow peak  as shown in Fig.~\ref{fig:transitions}.  

  The temperature at which the production rate is maximum, $T_{\rm max}$,  depends on $H(T)$, but it does not significantly change for the particular cosmologies we consider (see Fig.~\ref{fig:transitions}).  This can be seen in Eq.~\eqref{eq:mcmax} where $T_{\rm max}$ is given as a function of $\beta$ and $\epsilon$, for $\beta \leq 2$.  In the standard cosmology the production is maximal when the $V_T$ term is about 0.2 of the mass term (i.e. the square bracket in the denominator of Eq.~\eqref{eq:mattermixing} at $T_{\rm max}$ is 
$[1 - (2 \epsilon/m_s^2) T_{\rm max} V_T(T_{\rm max})] = [1 + 0.2]$) and is very similar in the other cosmologies. 
 
   For the standard cosmology ($\beta = 0$),
\begin{equation}
\label{eq:Stdmax}
    T_{\textrm{max}}^{\rm Std} = 145 \textrm{ MeV} \left(\frac{m_s}{\textrm{keV}}\right)^{\frac{1}{3}}\epsilon^{-\frac{1}{3}} \left(\frac{B}{10.88\times10^{-9}\textrm{ GeV}^{-4}}\right)^{-\frac{1}{6}}~.
\end{equation}
For ST1 ($\beta = -0.8$) $T_{\textrm{max}}$ is very similar
\begin{equation}\label{eq:st1tmax}
    T_{\textrm{max}}^{\rm ST1} = 156 \textrm{ MeV} \left(\frac{m_s}{\textrm{keV}}\right)^{\frac{1}{3}}\epsilon^{-\frac{1}{3}} \left(\frac{B}{10.88\times10^{-9}\textrm{ GeV}^{-4}}\right)^{-\frac{1}{6}}~.
\end{equation}
Since for ST1 the maximum temperature has an inverse scaling dependence on the momentum via $\epsilon$ (because $\beta$ is negative for this model, see Eq.~\eqref{eq:mcmax}), states with lower momentum will be produced earlier, at higher temperatures.

 We note that the usual definition of $T_{\rm max}$, as given by Dodelson and Widrow~\cite{Dodelson:1993je},  does not depend on $\epsilon$, because it is computed using the production rate integrated over momenta. It coincides with the definition of $T_{\rm max}$ given here for $\epsilon \simeq 1.3$.
 
 $T_{\rm max}^{\rm K}$ and $T_{\rm max}^{\rm ST2}$ for K and ST2 are given in  Appendix~\ref{ssecapp:maxtemp}. They are also very similar.

For non-standard cosmologies  with $\beta \geq 2$,  the sterile production rate $(\partial f_s/\partial T)_{\epsilon}$ is continuously increasing with decreasing $T$, as can be seen in Eq.~\eqref{eq:mcmax} for $\beta=2$.

\subsection{Sterile neutrino momentum distribution functions}

For the DW production mechanism, a closed-form expression for the momentum distribution function in a non-standard cosmology characterized by $H$ in Eq.~\eqref{eq:hnStd} can be found from Eq.~\eqref{eq:boltzmann2},
\begin{align}
\label{eq:mcdist}
    f_{\nu_s}(\epsilon) = \int_0^\infty \frac{A^{\prime} T^{2-\beta}}{(1+B'T^6)^2}\, f_{\nu_\alpha}dT = A^{\prime}\, B'^{(-\frac{1}{2}+\frac{\beta}{6})}\pi\, \Big(\frac{3+\beta}{36}\Big) \sec\left(\frac{\beta\pi}{6}\right)f_{\nu_\alpha}(\epsilon)~,
\end{align}
where $A'$ and $B'$ are
\begin{equation} 
\label{eq:apbp}
    A' = \eta^{-1}\sqrt{\dfrac{90}{8 g_{\ast} \pi^3}} M_{\rm Pl} \Gamma~, 
    ~~~~~~~~~~ B' = \dfrac{2B\epsilon^2}{m_s^2}~.
\end{equation}
Replacing in Eq.~\eqref{eq:mcdist} the expressions for $A'$ and $B'$ from Eq.~\eqref{eq:apbp} and $\Gamma$ from Eq.~\eqref{eq:interaction}, we obtain the distribution function for generic $\eta$ and $\beta$ given in Eq.~\eqref{eq:mcdist2} in the Appendix~\ref{ssecapp:distfunc}. In the Std cosmology $(\eta = 1, \beta = 0)$
\begin{align}
\label{eq:Stddist}
    f_{\nu_s}^{\rm Std}(\epsilon) = 9.25\times10^{-6}\left(\frac{\sin^2(2\theta)}{10^{-10}}\right)
    \left(\frac{m_s}{\textrm{keV}}\right) \left(\frac{g_{\ast}}{30}\right)^{-\frac{1}{2}} \left(\frac{d_\alpha}{1.13}\right) 
    \left(\frac{B}{10.88\times10^{-9}\textrm{GeV}^{-4}}\right)^{-\frac{1}{2}}f_{\nu_\alpha}(\epsilon)~.
\end{align}
For the ST1  cosmology ($\eta = 7.45 \times 10^{5}$, $\beta = -0.8$), the magnitude of the distribution is much smaller than in the Std cosmology,
\begin{align}
\label{eq:st1dist}
    f_{\nu_s}^{\rm ST1}(\epsilon) ~=&~ 1.96\times10^{-10} \times \epsilon^{-0.27} \left(\frac{\sin^2(2\theta)}{10^{-10}}\right) \left(\frac{m_s}{\textrm{keV}}\right)^{1.27}\left(\frac{T_{\rm tr}}{5 \textrm{ MeV}}\right)^{-0.82} \notag\\ 
    & \left(\frac{g_{\ast}}{30}\right)^{-\frac{1}{2}} \left(\frac{d_\alpha}{1.13}\right) \left(\frac{B}{10.88\times10^{-9}\textrm{ GeV}^{-4}}\right)^{-0.64} f_{\nu_\alpha}(\epsilon)~.
\end{align}
 The distribution functions for the K, ST2 and LRT cosmologies are given in the Appendix~\ref{ssecapp:distfunc}.
 
 We can clearly see from Eqs.~\eqref{eq:mcdist} and \eqref{eq:apbp}, and further in Eq.~\eqref{eq:mcdist2}), that cosmologies with larger expansion rates  (i.e.~larger $\eta$) require higher mixing angles to produce the same relic density, and vice-versa. For cosmologies with $\beta\neq0$, there is an extra momentum dependence that makes the distribution warmer (i.e. favors larger $\epsilon $ values) for $\beta>0$ or colder (favors smaller $\epsilon$ values) for $\beta<0$ than the standard Fermi-Dirac.
 
\subsection{Sterile neutrino number densities}
\label{ssec:numberdensity}

\begin{figure}[htb]
\begin{center}
\includegraphics[trim={5mm 0mm 40 0},clip,width=.475\textwidth]{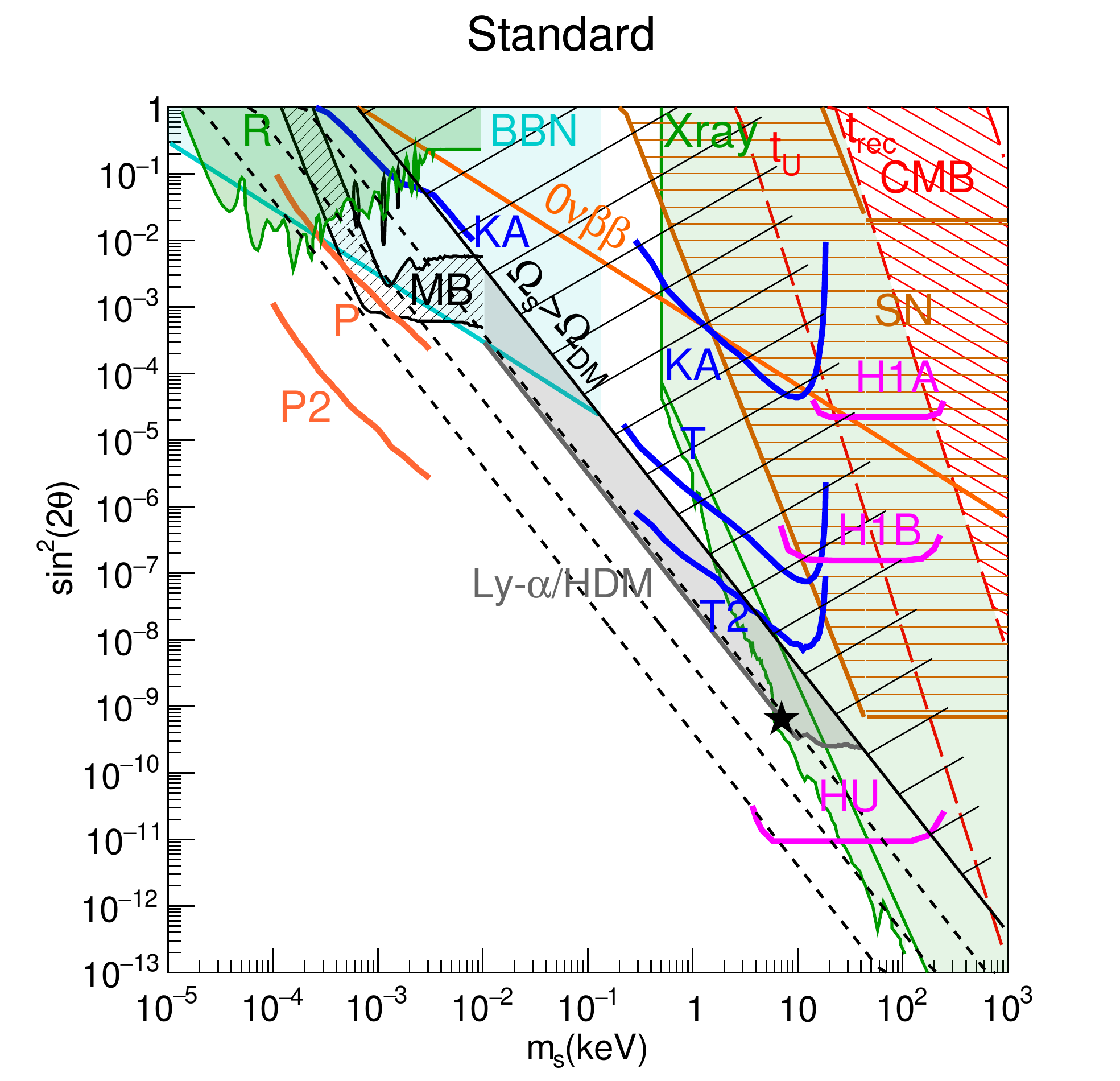}
\includegraphics[trim={5mm 0mm 40 0},clip,width=.475\textwidth]{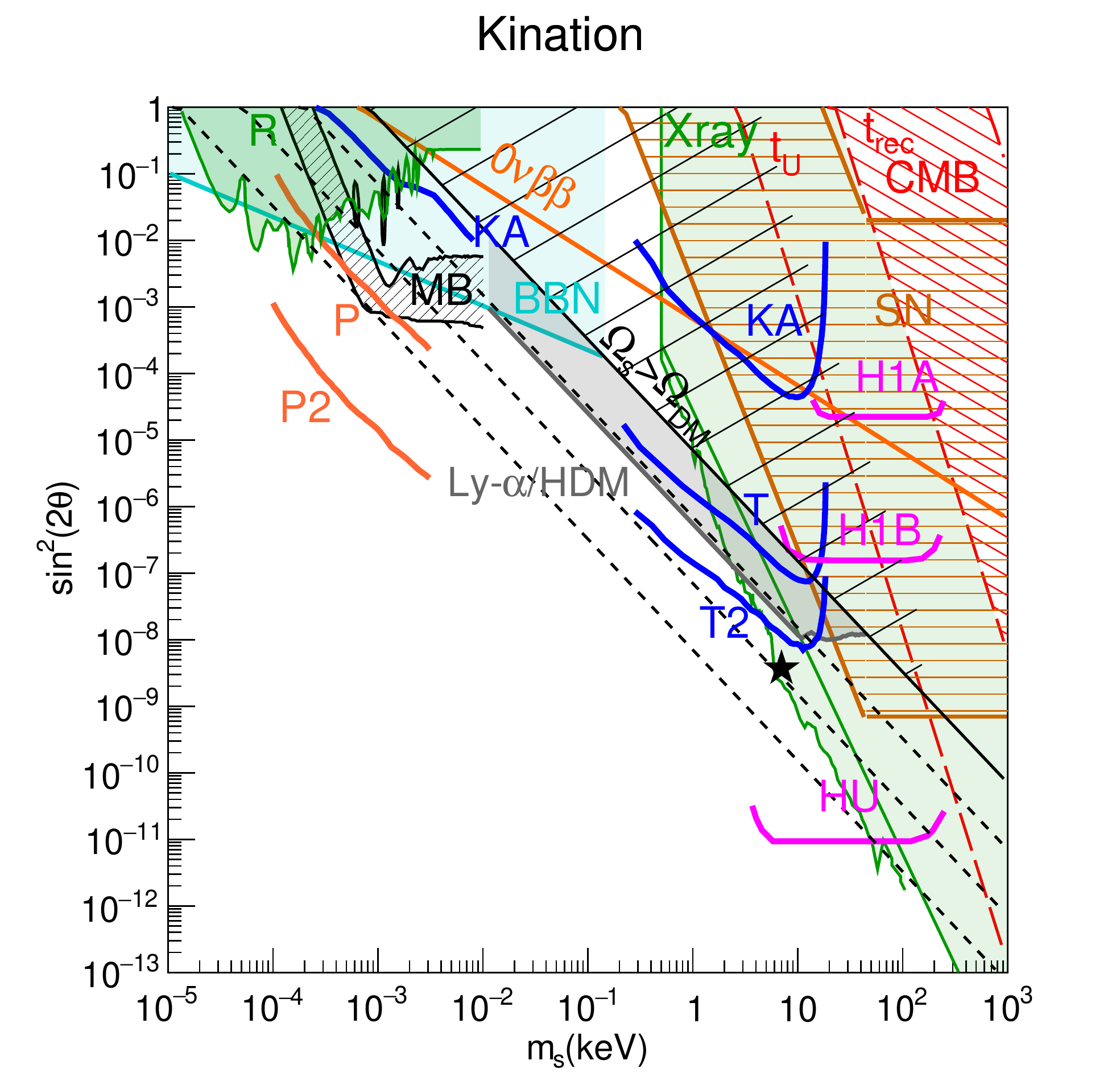}
\includegraphics[trim={5mm 0mm 40 0},clip,width=.475\textwidth]{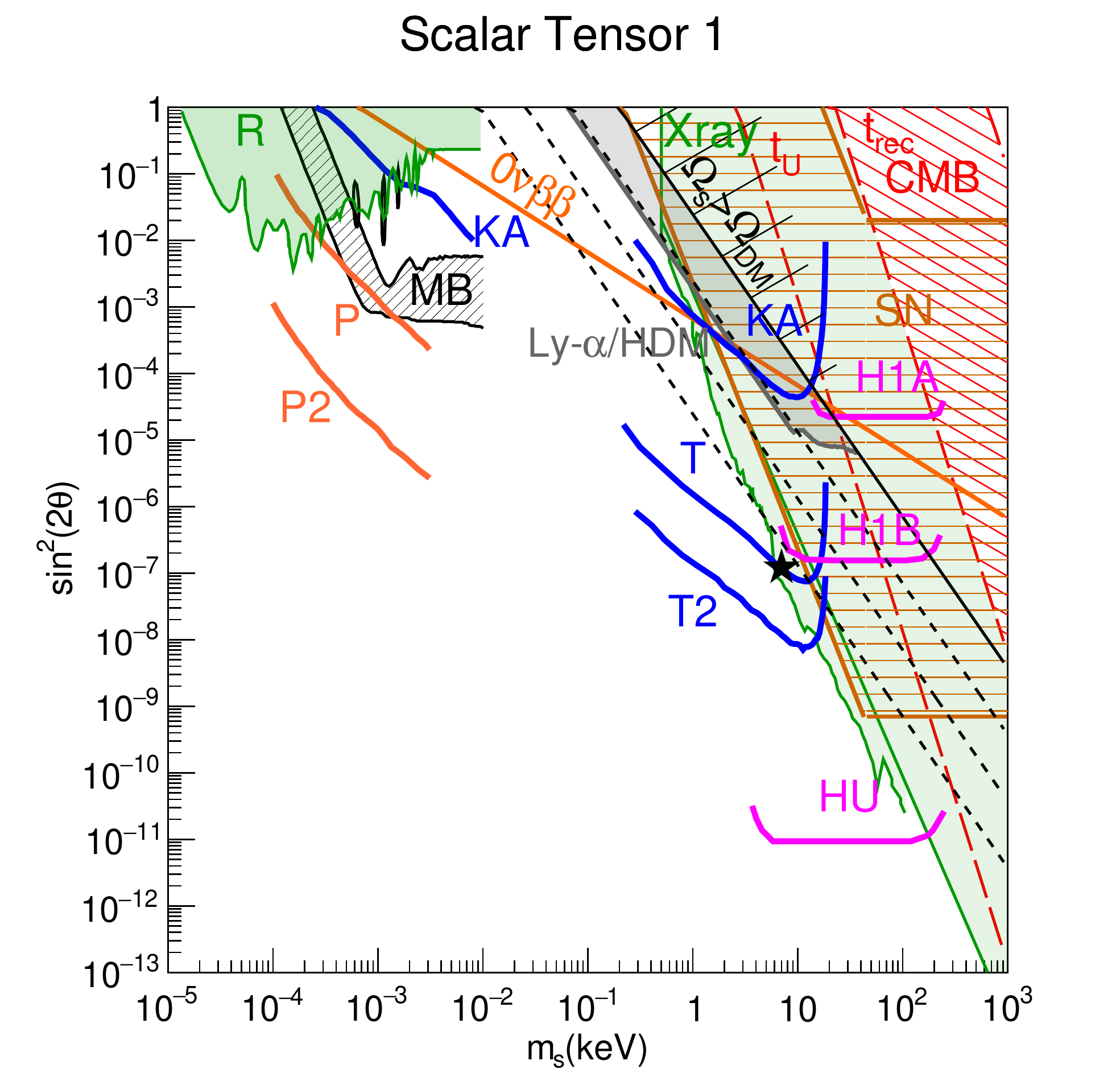}
\includegraphics[trim={5mm 0mm 40 0},clip,width=.475\textwidth]{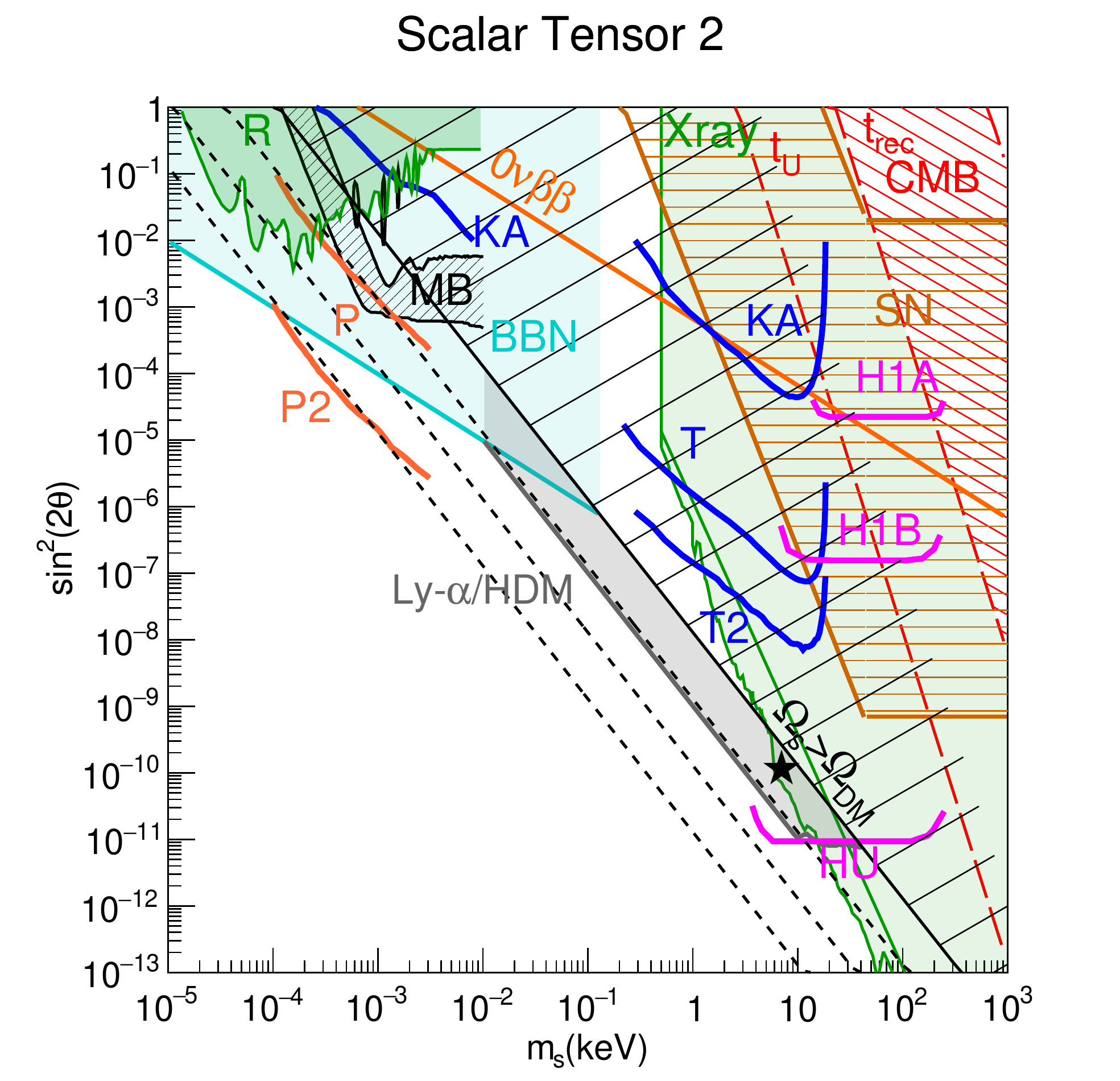}
\caption{  \label{fig:allDWlim}  Present relic abundance, limits and regions of interest for standard, kination and scalar-tensor cosmologies. See caption in  Fig.~\ref{fig:lrDWlim}.}
\end{center}
\end{figure}

The number density $n_{\nu_\alpha}$ of active neutrinos is obtained by integration over momentum of the distribution function of Eq.~\eqref{eq:fermid} 
\begin{equation}
\label{eq:activenumden}
    n_{\nu_{\alpha}}(T_{\nu_\alpha}) = \left(\frac{3\zeta(3)}{2\pi^2}\right) T_{\nu_\alpha}^3,
\end{equation}
where for photon temperatures $T>1$ MeV, the temperature of the neutrino background is $T_{\nu_\alpha}= T$, and for $T<1$ MeV, $T_{\nu_\alpha}= (4/11) T$.  The number density $n_{\nu_s}$ of sterile neutrinos can be obtained in a similar manner
\begin{equation}
\label{eq:distintegrate}
    n_{\nu_s}(T_{\nu_s}) = 2\int^\infty_0 \frac{d^3 p}{(2\pi)^3} f_{\nu_s}(p) = \frac{T_{\nu_s}^3}{\pi^2} \int^\infty_0 d \epsilon ~ \epsilon^2 f_{\nu_s}(\epsilon)= T_{\nu_s}^3 C F_{2+\frac{\beta}{3}}(0)~,
\end{equation}
considering that active neutrinos  have a Fermi-Dirac distribution, as given in Eq.~\eqref{eq:fermid},  with $\xi = 0$ (since for non-resonant production the chemical potential of active neutrinos is negligible)   and $T_{\nu_s}$ being the temperature of the sterile neutrino background. Here, 
 $C = f_{\nu_s}(\epsilon)\left(\pi^2 \epsilon^{\beta/3} f_{\nu_\alpha} (\epsilon)\right)^{-1}$ is a constant (notice that in Eq.~\eqref{eq:mcdist2}  $f_{\nu_s}$ depends on $\epsilon$ only through the product $\epsilon^{\beta/3} f_{\nu_\alpha}(\epsilon)$) and 
\begin{equation}
\label{eq:relfermi}
F_k(\xi) = \int_{0}^{\infty} dx \dfrac{x^k}{e^{x - \xi} + 1}~,    
\end{equation}
is the relativistic Fermi integral.  The relic sterile neutrino number density as function of $\eta$ and $\beta$ is given in  Eq.~\eqref{eq:numden}. 

\begin{figure}[htb]
\begin{center}
\includegraphics[trim={5mm 0mm 40 0},clip,width=.475\textwidth]{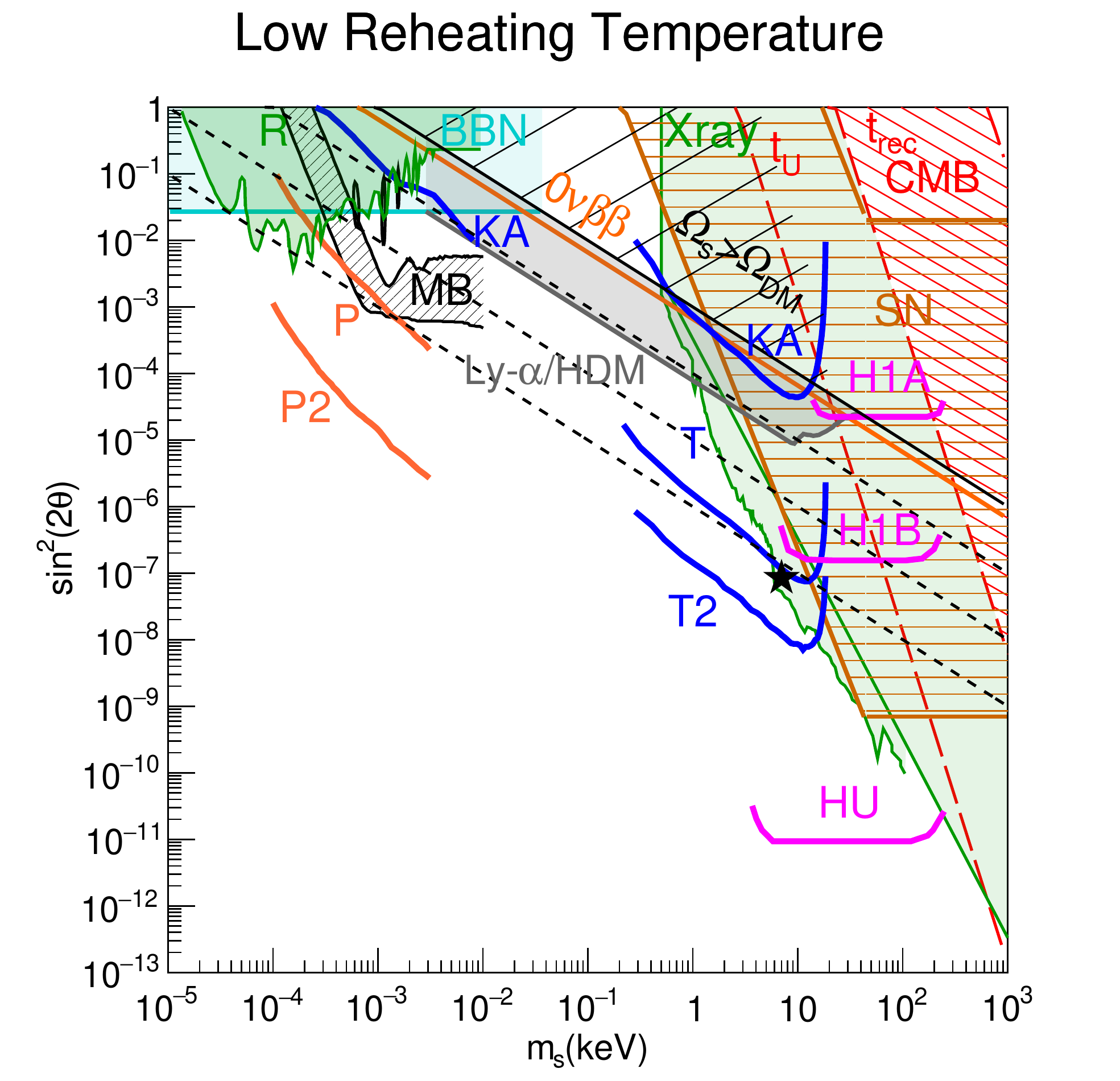}
\caption{\label{fig:lrDWlim} Present relic abundance, limits and regions of interest in the mass-mixing space of a  $\nu_s$ mixed with $\nu_e$, for LRT cosmology with $T_{\rm RH} = 5$ MeV~\cite{Gelmini:2004ah}. Shown are the fraction of the DM in $\nu_s$ of 1 (black solid line) and $10^{-1}$, $10^{-2}$ and $10^{-3}$ (black dotted lines), the forbidden region $\Omega_s/\Omega_{\rm DM} > 1$  (diagonally hatched in black), lifetimes $\tau= t_U$, $t_{\rm rec}$ and $t_{\rm th}$ (see text) of Majorana $\nu_s$ (straight long dashed red lines), the region (SN) disfavored by supernovae~\cite{Kainulainen:1990bn} (horizontally hatched in brown), the location  of the 3.5 keV X-ray signal~\cite{Bulbul:2014sua,Boyarsky:2014jta} for each cosmology (black star). The regions rejected by reactor neutrino (R) experiments (Daya Bay~\cite{An:2016luf}, Bugey-3~\cite{Declais:1994su} and  PROSPECT~\cite{Ashenfelter:2018iov}) shown in green, limits on $N_{\rm eff}$ during BBN~\cite{Tanabashi:2018oca} (BBN) in cyan,  Lyman-alpha limits~\cite{Baur:2017stq} (Ly-$\alpha$/HDM) in gray, X-ray limits~\cite{Ng:2019gch,Perez:2016tcq,Neronov:2016wdd} including DEBRA~\cite{Boyarsky:2005us}  (Xray) in green, $0\nu\beta\beta$ decays~\cite{KamLAND-Zen:2016pfg}  ($0\nu\beta\beta$) in orange  and CMB spectrum distortions~\cite{Fixsen:1996nj} (CMB) diagonally hatched in red. Current/future sensitivity of KATRIN (KA) in the keV~\cite{Mertens:2018vuu} and  eV~\cite{megas:thesis}  mass range, its TRISTAN upgrade in 1 yr (T) and in 3 yr (T2)~\cite{Mertens:2018vuu} shown by blue solid lines. Magenta solid lines  show the reach of the  phases 1A (H1A) and 1B (H1B) of HUNTER, and its upgrade (HU)~\cite{Smith:2016vku}.  The 4-$\sigma$ band of compatibility with LSND and MiniBooNE results (MB) in Fig.~4 of~\cite{Aguilar-Arevalo:2018gpe} is shown densely hatched in black.  The three black vertical elliptical  contours are the regions allowed at 3-$\sigma$ by DANSS~\cite{Alekseev:2018efk} and NEOS~\cite{Ko:2016owz} data in Fig.~4 of~\cite{Gariazzo:2018mwd}). Orange solid lines show the reach of PTOLEMY  for 10 mg-yr (P) and 100 g-yr (P2) exposures (from Figs. 6 and 7 of~\cite{Betti:2019ouf}).}
\end{center}
\end{figure}

The term $\beta/3$ in the definition of the index $k=2+ \beta/3$ comes from the $\epsilon^{\beta/3}$ dependence of $f_{\nu_s}$ in  Eq.~\eqref{eq:mcdist2} (for LRT, the index is instead $k=3$ because of the $\epsilon^1$ dependence of $f_{\nu_s}^{\textrm{LRT}}$ in  Eq.~\eqref{eq:lrtdist}).  

The ratio of sterile and active neutrino number densities at the same temperature can be easily stated in terms of the ratio of momentum distributions. For the parametrization of $H$ in Eq.~\eqref{eq:hnStd}, this ratio is given in Eq.~\eqref{eq:numratio-gen}.

In the standard cosmology, for which ${n_{\nu_s}(T)}/{n_{\nu_a}(T)} =  {f_{\nu_s}(\epsilon)}/{f_{\nu_\alpha}(\epsilon)}$, the  relic sterile neutrino number density is
\begin{align}
\label{eq:Stdnumden-general}
    n_{\nu_s}^{\rm Std}(T_{\nu_s})~=&~ 9.25\times10^{-6} \left(\frac{\sin^2 2\theta}{10^{-10}}\right)\left(\frac{m_s}{\textrm{keV}}\right)\left(\frac{T_{\nu_s}}{T_{\nu_\alpha}}\right)^3\left(\frac{g_\ast}{30}\right)^{-\frac{1}{2}} \notag \\ & \times\left(\frac{d_\alpha}{1.13}\right)\left(\frac{B}{10.88\times10^{-9}\textrm{GeV}^{-4}}\right)^{-\frac{1}{2}} n_{\nu_\alpha}(T_{\nu_\alpha})~.
\end{align}
Thus, the present number density is 
\begin{align}
\label{eq:Stdnumden}
    n_{\nu_s}^{\rm Std}~=&~3.74\times10^{-4}\textrm{cm}^{-3}\left(\frac{\sin^2 2\theta}{10^{-10}}\right)\left(\frac{m_s}{\textrm{keV}}\right)\left(\frac{T_{\nu,0}}{1.95~{\rm K}}\right)^3 \notag\\ &\times \left(\frac{g_{\ast}}{30}\right)^{-\frac{3}{2}}\left(\frac{d_\alpha}{1.13}\right) \left(\frac{B}{10.88\times10^{-9}\textrm{ GeV}^{-4}}\right)^{-\frac{1}{2}}~,
\end{align}
where $T_{\nu,0} = 1.95~{\rm K} = 0.17$ meV is the present temperature of the active relic neutrino background. 

In the ST1 model, ${n_{\nu_s}(T)}/{n_{\nu_a}(T)} = 0.77~ \epsilon^{0.27} ({f_{\nu_s}(\epsilon)}/{f_{\nu_\alpha}(\epsilon)})$, and the 
number density is significantly reduced for the same mass and mixing angle,
\begin{align}
\label{eq:st1numden-general}
    n_{\nu_s}^{\rm ST1}(T_{\nu_\alpha}) ~=&~ 1.52\times10^{-10}\left(\frac{\sin^2 2\theta}{10^{-10}}\right)\left(\frac{m_s}{\textrm{keV}}\right)^{1.27}\left(\frac{T_{\rm tr}}{5\textrm{ MeV}}\right)^{-0.82}\left(\frac{T_{\nu_s}}{T_{\nu_\alpha}}\right)^3 \notag\\
    &\times\left(\frac{g_{\ast}}{30}\right)^{-\frac{1}{2}}\left(\frac{d_\alpha}{1.13}\right)  \left(\frac{B}{10.88\times10^{-9}\textrm{ GeV}^{-4}}\right)^{-0.64} n_{\nu_\alpha} (T_{\nu_\alpha})~,
\end{align}
with the present number density being
\begin{align}
\label{eq:st1numden}
    n_{\nu_s}^{\rm ST1}=&~ 6.13\times10^{-9}~\textrm{cm}^{-3}\left(\frac{\sin^2 2\theta}{10^{-10}}\right)\left(\frac{m_s}{\textrm{keV}}\right)^{1.27}\left(\frac{T_{\rm tr}}{5\textrm{ MeV}}\right)^{-0.82}\left(\frac{T_{\nu,0}}{1.95~{\rm K}}\right)^3 \notag\\ 
    &\times\left(\frac{g_{\ast}}{30}\right)^{-\frac{3}{2}}\left(\frac{d_\alpha}{1.13}\right)  \left(\frac{B}{10.88\times10^{-9}\textrm{ GeV}^{-4}}\right)^{-0.64}~.
\end{align}
Here, $g_{\ast}$ is the number of relativistic degrees of freedom when sterile neutrinos are produced, which we take to be $g_{\ast}(T_{\rm max})$. Thus, at present, the ratio of temperatures of the sterile and active neutrinos is 
$(T_{\nu_s,0}/T_{\nu, 0}) = (10.75/g_{\ast})^{1/3}$. 

The present number densities for the K and ST2 and also LRT cosmologies are given in the Appendix~\ref{ssecapp:numden}.

\subsection{Relativistic energy density}
\label{ssec:relenergy}

The energy density of relativistic active neutrinos is given by (Eq. \eqref{eq:fermid})
\begin{equation}
\label{eq:activeenergyden}
    \rho_{\nu_\alpha} = 2\int^\infty_0 \frac{d^3 p}{(2\pi)^3} ~p~f_{\nu_\alpha}(p) = \frac{T^4}{\pi^2} \int^\infty_0 d \epsilon ~ \epsilon^3 f_{\nu_\alpha}(\epsilon)= \frac{T^4}{\pi^2} F_{3}(0)~,
\end{equation}
where the Fermi integral $F_k (\xi)$ is defined in Eq.~\eqref{eq:relfermi}. Similarly, the energy density of relativistic  sterile neutrinos is 
\begin{equation}
\label{eq:distintegrate2}
    \rho_{\nu_s} = 2\int^\infty_0 \frac{d^3 p}{(2\pi)^3} ~p~f_{\nu_s}(p) = \frac{T^4}{\pi^2} \int^\infty_0 d \epsilon ~ \epsilon^3 f_{\nu_s}(\epsilon)= T^4 C F_{3+\frac{\beta}{3}}(0)~.
\end{equation}
To obtain the last equality we used the momentum distributions for the models with $H$  given in Eq.~\eqref{eq:hnStd}.  The resulting energy density in terms of the parameters $\eta$ and $\beta$ is given in Eq.~\eqref{eq:genrhonus}. 

For the LRT model, the index in the function $F_k$ in this equation  is not $k= 3+ \beta/3$ but instead $k=4$ because of the $\epsilon$ dependence of $f_{\nu_s}^{\textrm{LRT}}$ in  Eq.~\eqref{eq:lrtdist}.  

The ratio of sterile and active neutrino relic densities at the same temperature $T$ can be easily stated in terms of the ratio of momentum distributions. This ratio is given in terms of $\eta$ and $\beta$ in Eq.~\eqref{eq:densratio-gen}.

In the Std cosmology, ${\rho_{\nu_s}(T)}/{\rho_{\nu_a}(T)}= {n_{\nu_s}(T)}/{n_{\nu_a}(T)} =  {f_{\nu_s}(\epsilon)}/{f_{\nu_\alpha}(\epsilon)}$, and the energy density of non-resonantly produced relativistic sterile neutrinos with temperature $T_{\nu_s}$ is 
\begin{align}
\label{eq:Stdrelden}
    \rho_{\nu_s}^{\rm Std}(T_{\nu_s}) ~=&~ 9.25\times10^{-6} ~~ \left(\frac{\sin^2 2\theta}{10^{-10}}\right) \left(\frac{m_s}{\textrm{keV}}\right)\left(\frac{T_{\nu_s}}{T_{\nu_\alpha}}\right)^4\left(\frac{g_{\ast}}{30}\right)^{-\frac{1}{2}} \notag\\ 
    &\times  \left(\frac{d_\alpha}{1.13}\right) \left(\frac{B}{10.88\times10^{-9}\textrm{ GeV}^{-4}}\right)^{-\frac{1}{2}} \rho_{\nu_\alpha}(T_{\nu_\alpha})~,
\end{align}
or,
\begin{align}
\label{eq:Stdrelden-2}
    \rho_{\nu_s}^{\rm Std}(T_{\nu_s}) 
    =&~ 6.93\times10^{26} ~~ \dfrac{\textrm{MeV}}{\textrm{cm}^{3}}\left(\frac{\sin^2 2\theta}{10^{-10}}\right) \left(\frac{m_s}{\textrm{keV}}\right)\left(\frac{T_{\nu_s}}{1~{\rm MeV}}\right)^4\left(\frac{g_{\ast}}{30}\right)^{-\frac{1}{2}} \notag\\ 
    &\times  \left(\frac{d_\alpha}{1.13}\right) \left(\frac{B}{10.88\times10^{-9}\textrm{ GeV}^{-4}}\right)^{-\frac{1}{2}}~.
\end{align}
For ST1 instead, ${\rho_{\nu_s}(T)}/{\rho_{\nu_a}(T)}= 0.92 ({n_{\nu_s}(T)}/{n_{\nu_a}(T)})$ and the density is much smaller
 \begin{align}
\label{eq:st1relden}
    \rho_{\nu_s}^{\rm ST1}(T_{\nu_s}) ~=&~ 1.39\times10^{-10}~\left(\frac{\sin^2 2\theta}{10^{-10}}\right)\left(\frac{m_s}{\textrm{keV}}\right)^{1.27} \left(\frac{T_{\rm tr}}{5\textrm{ MeV}}\right)^{-0.82}\left(\frac{T_{\nu_s}}{T_{\nu_\alpha}}\right)^4 \left(\frac{g_{\ast}}{30}\right)^{-\frac{1}{2}}
     \notag\\ 
    &\times  \left(\frac{d_\alpha}{1.13}\right) \left(\frac{B}{10.88\times10^{-9}\textrm{ GeV}^{-4}}\right)^{-0.64} \rho_{\nu_\alpha}(T_{\nu_\alpha}) ~,
\end{align}
or,
 \begin{align}
\label{eq:st1relden}
    \rho_{\nu_s}^{\rm ST1}(T_{\nu_s})
    =&~ 1.04\times10^{22}~\dfrac{\textrm{MeV}}{\textrm{cm}^{3}}\left(\frac{\sin^2 2\theta}{10^{-10}}\right)\left(\frac{m_s}{\textrm{keV}}\right)^{1.27} \left(\frac{T_{\rm tr}}{5\textrm{ MeV}}\right)^{-0.82}\left(\frac{T_{\nu_s}}{1~{\rm MeV}}\right)^4 \left(\frac{g_{\ast}}{30}\right)^{-\frac{1}{2}}
     \notag\\ 
    &\times  \left(\frac{d_\alpha}{1.13}\right) \left(\frac{B}{10.88\times10^{-9}\textrm{ GeV}^{-4}}\right)^{-0.64}~.
\end{align}
 The relativistic energy density $\rho_{\nu_s}^{\rm K}$, $\rho_{\nu_s}^{\rm ST2}$ and $\rho_{\nu_s}^{\rm LRT}$ for the  K, ST2 and LRT cosmologies are given in Appendix~\ref{ssecapp:relenergy}. 
 
 These expressions apply after the bulk of  the sterile neutrinos has been produced and while they are relativistic: $m_s < T < T_\textrm{max}$ (or for $m_s < T < T_\textrm{RH}$ in LRT models).

The average $T$-scaled dimensionless sterile neutrino  momentum $\langle \epsilon \rangle$ in the different cosmologies is given in terms of the parameters $\eta$ and $\beta$ in Eq.~\eqref{eq:hnStd} for $T_{\rm tr}=$ 5 MeV is
\begin{equation}
\label{eq:nonreseps}
\langle\epsilon\rangle = \dfrac{\rho_{\nu_s}(T)}{ T~ n_{\nu_s}(T)} =  \dfrac{F_{3+\beta/3}(0)}{F_{2+\beta/3}(0)} 
= \left \{
  \begin{tabular}{cl}
  3.15, & ~~~\text{Std}\\
  3.47, & ~~~\text{K} \\
  2.89, & ~~~\text{ST1} \\
  3.15, & ~~~\text{ST2} 
  \end{tabular}
\right .\
\end{equation}
where $F_k (\xi)$ is given in Eq.~\eqref{eq:relfermi} as before. For the LTR model with $T_{\rm RH}=$ 5 MeV
\begin{equation}
\label{eq:nonresepsLRT}
\langle\epsilon\rangle = \dfrac{\rho_{\nu_s}(T)}{ T~ n_{\nu_s}(T)} =  \dfrac{F_{4}(0)}{F_{3}(0)} 
= ~~~~~~~ 4.11,  ~~~~~\text{LRT}
\end{equation}

\subsection{Present fraction of the DM in non-resonantly produced sterile neutrinos}
\label{ssec:relicdensity}

The present sterile neutrino relic density must not exceed the DM density, i.e.  $\Omega_{\nu_s} = \rho_{\nu_s}/\rho_{\rm crit} \leq \Omega_{\rm DM}=\rho_{\rm DM}/\rho_{\rm crit}$,  where  $\Omega_{DM} h^2 = 0.1186 \simeq 0.12$,  $\rho_{\rm crit} = 1.054 \times 10^{-5} h^2$ GeV/cm$^{3}$ is the critical density of the Universe and $h = 0.678$~\cite{Tanabashi:2018oca}. Hence, the fraction of the DM consisting of sterile neutrinos $f_{s, {\rm DM}}=\rho_{\nu_s}/\rho_{\rm DM}$  must be $f_{s, {\rm DM}} \leq 1$. 

At present, all the sterile neutrinos we consider (i.e.~all neutrinos with $m_s >10^{-2}$ eV), are non-relativistic,  thus $\rho_{\nu_s,0} = m_s n_{\nu_s,0}$. Hence, the present fraction of the DM in sterile neutrinos is $f_{s, {\rm DM}}= (n_{\nu_s,0}~  m_s/ \rho_{\textrm{DM}})$.  This fraction is given as a function of the $\eta$ and $\beta$ parameters in Eq.~\eqref{eq:overden}.  In the Std cosmology  this fraction is
\begin{align}
\label{eq:Stdoverden}
        f_{s, {\rm DM}}^{\rm Std} =&~\Big(\dfrac{n_{\nu_s,0} \, m_s}{\rho_{\textrm{DM}}}\Big)^{\rm Std}= \notag\\ ~=&~ 2.48\times10^{-4}\left(\frac{\sin^2(2\theta)}{10^{-10}}\right) \left(\frac{m_s}{\textrm{keV}}\right)^{2}\left(\frac{T_{\nu, 0}}{1.95~{\rm K}}\right)^3  \notag\\ & \times \left(\frac{g_{\ast}}{30}\right)^{-\frac{3}{2}}\left(\frac{d_\alpha}{1.13}\right)\left(\frac{\Omega_\textrm{DM}h^2}{ 0.12}\right)\left(\frac{B}{10.88\times10^{-9}\textrm{ GeV}^{-4}}\right)^{-\frac{1}{2}}
\end{align}
and in the ST1 cosmology is instead
\begin{align}
\label{eq:st1overden}
       f_{s, {\rm DM}}^{\rm ST1} =&~    \Big(\dfrac{n_{\nu_s,0} \, m_s}{\rho_{\textrm{DM}}}\Big)^{\rm ST1} = \notag\\ ~=&~ 4.09\times10^{-9}\left(\frac{\sin^2(2\theta)}{10^{-10}}\right)\left(\frac{m_s}{\textrm{keV}}\right)^{2.27} \left(\frac{T_{\rm tr}}{5\textrm{ MeV}}\right)^{-0.82}\left(\frac{T_{\nu,0}}{1.95~{\rm K}}\right)^3 \notag\\ & \times\left(\frac{g_{\ast}}{30}\right)^{-\frac{3}{2}}\left(\frac{d_\alpha}{1.13}\right) \left(\frac{\Omega_\textrm{DM}h^2}{ 0.12}\right)\left(\frac{B}{10.88\times10^{-9}\textrm{ GeV}^{-4}}\right)^{-0.64}~.
\end{align}
The present  sterile neutrino fraction of the DM for 
K, ST2  and also the LRT cosmologies are given in Appendix~\ref{ssecapp:abundance}.
 
 The condition $f_{s, {\rm DM}} = 1$  defines  the mixing  we call $\sin^2(2\theta)_\textrm{DW,lim}$ as a  function of  $m_s$.  In the  log-log scales used in our figures,  this is a straight line on which sterile neutrinos account for the entirety of the DM. As a function of $\eta$ and $\beta$, $\sin^2(2\theta)_\textrm{DW,lim}(m_s)$ is given in Eq.~\eqref{eq:overdenselimit}. 
 
 In the Std cosmology
\begin{align}
\label{eq:Stdoverdenselimit}
    \sin^2(2\theta)_\textrm{DW,lim}^{\rm STD} ~=&~ 4.02\times10^{-7}\left(\frac{m_s}{\textrm{keV}}\right)^{-2}  \left(\frac{T_{\nu, 0}}{1.95~{\rm K}}\right)^{-3} \left(\frac{g_{\ast}}{30}\right)^{\frac{3}{2}}\left(\frac{d_\alpha}{1.13}\right)^{-1}\notag\\ &\times \left(\frac{\Omega_\textrm{DM}h^2}{ 0.12}\right)^{-1}\left(\frac{B}{10.88\times10^{-9}\textrm{ GeV}^{-4}}\right)^{\frac{1}{2}} 
\end{align}
and in the  ST1 cosmology
\begin{align}
\label{eq:st1overdenselimit}
    \sin^2(2\theta)_\textrm{DW,lim}^{\rm ST1} ~=&~ 2.46\times10^{-2}\left(\frac{m_s}{\textrm{keV}}\right)^{-2.27} \left(\frac{T_{\rm tr}}{5 \textrm{ MeV}}\right)^{0.82} \left(\frac{T_{\nu, 0}}{1.95~{\rm K}}\right)^{-3} \left(\frac{g_{\ast}}{30}\right)^{\frac{3}{2}}\notag\\ &\times \left(\frac{d_\alpha}{1.13}\right)^{-1}\left(\frac{\Omega_\textrm{DM}h^2}{ 0.12}\right)^{-1}\left(\frac{B}{10.88\times10^{-9}\textrm{ GeV}^{-4}}\right)^{0.64} ~.
\end{align}
In Appendix~\ref{ssecapp:overdensity} this function is given for the K and ST2 cosmologies, and also for the LRT (see Eq.~\eqref{eq:lrtoverdenselimit}).

In Figs.~\ref{fig:allDWlim} and \ref{fig:lrDWlim}, the fractions $f_{s, {\rm DM}}=$ 1 is indicated with a solid black lines and $f_{s, {\rm DM}}= 10^{-1}, 10^{-2}, 10^{-3}$ are indicated with dotted black lines.

\section{Limits and potential signals}
\label{sec:limits}
 
Here we discuss constraints, regions of interest  in the mass-mixing plane and potential signals  for  sterile neutrinos produced via non-resonant active-sterile oscillations in different pre-BBN cosmologies. The results are shown in Figs.~\ref{fig:allDWlim} and~\ref{fig:lrDWlim}.
In the figures we assume that the sterile neutrino only mixes with $\nu_e$.

The most stringent  cosmological and astrophysical limits on sterile neutrinos  come from Lyman-$\alpha$ forest  and X-ray  observations for the 1 to 10 keV mass range and BBN for $m_s < 100$ eV. We also discuss the 3.5 keV X-ray line as well as upcoming laboratory experiments such as KATRIN/TRISTAN~\cite{Mertens:2018vuu} and HUNTER~\cite{Smith:2016vku} for the 1 to 10 keV mass range, reactor and accelerator neutrino experiments  for $m_s < 10$ eV and also the $0\nu \beta \beta$ searches.

\subsection{\texorpdfstring{Lyman-$\alpha$}{lalpha} forest WDM and HDM limits}
\label{ssec:lyman}

Sterile neutrinos of $m_s= \mathcal{O}$(keV) produced non-resonantly constitute a warm DM (WDM) candidate. The free-streaming of DM particles suppresses structure formation below the free-streaming scale (see e.g. Ref.~\cite{Boyarsky:2018tvu})
\begin{equation} 
\label{eq:freestream}
\lambda_{\rm fs} = a(t_U) \int_{t_i}^{t_U} dt^{\prime} \dfrac{v(t^{\prime})}{a(t^{\prime})} \simeq 1~\text{Mpc} \Big(\dfrac{\text{keV}}{m_s}\Big)  \dfrac{\langle p_{\nu_s}\rangle}{\langle p_{\nu_{\alpha}}\rangle} = 1~\text{Mpc}   \Big(\dfrac{\langle \epsilon \rangle}{3.15}\Big)\Big(\dfrac{\text{keV}}{m_s}\Big)\dfrac{T_{\nu_s}}{ T_{\nu,0}} ~,   
\end{equation}
where $v(t)$ is a typical DM velocity,  $t_U$ is the present lifetime of the Universe, $t_i$ is some initial very early time whose exact value is not important, $a(t)$ is the scale factor of the Universe, $ \langle p_{\nu_s} \rangle$ and $\langle p_{\nu_{\alpha}}\rangle$ are the average absolute values of the sterile and active neutrino momentum and $T_{\nu_s}$ and $T_{\nu,0}$ are the present temperatures of the sterile and active neutrinos, respectively. 

Observations of the Lyman-$\alpha$ forest\footnote{An alternative approach is to use DM halo counts, whose number and formation are also related to free-streaming scales (see  the discussion in Ref.~\cite{Boyarsky:2018tvu}).}  in the spectra of distant quasars constrain the power spectrum on $\sim 0.1-1$ Mpc scales~\cite{Baur:2017stq}, which from Eq.~\eqref{eq:freestream} constrains WDM and thus sterile neutrinos with $m_s = \mathcal{O}$(keV).

The Lyman-$\alpha$ limits are usually given in terms of the mass $m_{\rm therm}$ of a fermion that was in thermal equilibrium at some point in the history of the Universe, and thus has a  Fermi-Dirac spectrum, and thus has $\langle\epsilon\rangle =3.15$,  and a temperature $T_{\rm therm}$ that depends  on when the fermion decoupled from the radiation bath. Hence, the relic density $\Omega_{\rm therm}$ of this thermal fermion, which depends on its temperature and mass, and its mass are free independent parameters. Sterile neutrinos produced via active-sterile oscillations do not have a thermal equilibrium spectrum, and could have $\langle\epsilon\rangle \not=$ 3.15 (see Eqs.~\eqref{eq:nonreseps} and \eqref{eq:nonresepsLRT}). Sterile neutrinos may also have a temperature $T_{\nu_s}$ smaller than the active neutrino temperature $T_\nu$.  In fact, for  $m_s = \mathcal{O}$(keV) the temperature of maximum production $T_{\rm max} > 100$ MeV (except in the LRT models) is  much higher than the active neutrino decoupling temperature $T \simeq 3$ MeV,  resulting in  $T_{\nu_s} <T_{\nu}$.

 Following Ref.~\cite{Viel:2005qj}, we equate the free-streaming scales of thermal WDM  and sterile neutrinos
\begin{equation}
\dfrac{\langle\epsilon\rangle~ T_{\nu_s}}{m_s}  =\dfrac{3.15 ~T_{\rm therm}}{m_{\rm therm}}~,
\end{equation}
 as well as their energy densities,  $\Omega_{\rm therm}= \Omega _{\nu_s}=f_{\rm s,DM}~\Omega_{DM}$. This allows to identify the sterile neutrino mass that would result in the same free-streaming as a thermal  WDM fermion with  mass $m_{\rm therm}$
\begin{equation}
\label{eq:wdmdict-1}
m_s = 4.46 ~{\rm keV}~\left(\frac{\langle\epsilon\rangle}{3.15}\right)~
\left(\frac{m_\textrm{therm}}{\textrm{keV}}\right)^{\frac{4}{3}}~
\left(\dfrac{T_{\nu_s}}{T_{\nu_\alpha}}\right)~\left(\dfrac{0.12}{f_{\rm s,DM}~\Omega_{DM}~ h^2}\right)^{\frac{1}{3}}~,
\end{equation}
where  $T_{\nu_s}/T_{\nu_\alpha} = (10.75/g_{\ast})^{1/3}$. For our figures we use $g_{\ast}=30$ except for the LRT cosmology for which we take $g_{\ast}=15$.

The Lyman-$\alpha$ bounds of Ref.~\cite{Baur:2017stq} on   $m_{\rm therm}$  can now be translated into bounds on $m_s$ through Eq.~\eqref{eq:wdmdict-1}. We use the $2$-$\sigma$  bounds from the right panel of  Fig.~6 of Ref.~\cite{Baur:2017stq} coming from  SDSS, XQ  and HR data.  These limits impose that sterile neutrinos do not account for more than $\sim8\%$ of the DM density for $m_s \lesssim 1$ keV. Lower $m_s$ values  yield larger free-streaming lengths, which for $\lambda_{\rm fs} \gtrsim 1$ Mpc scales will not significantly alter the Lyman-$\alpha$ bounds. 
Hence,  in Figs.~\ref{fig:allDWlim} and \ref{fig:lrDWlim} we extend  the saturated bound of Ref.~\cite{Baur:2017stq}  to lighter sterile neutrinos that would constitute Hot DM (HDM), until it is superseded by the BBN bound in Eq.~\eqref{eq:DWNeff}. 

\subsection{BBN limit on the effective number of neutrino species}
\label{ssec:neff}

The impact on BBN of an increased expansion rate of the Universe  yields  an upper limit on  $N_{\rm eff}$, the effective number of relativistic active neutrino species present during BBN. Assuming that only sterile neutrinos and SM active neutrinos contribute  to $N_{\rm eff}$,
\begin{equation} 
N_{\rm eff} = 3.045+ \Big(\dfrac{\rho_{\nu_s}}{\rho_{\nu_\alpha}}\Big)~,
\end{equation}
where $3.045$ is the contribution of the SM active neutrinos alone~\cite{Mangano:2005cc, deSalas:2016ztq}. All the sterile neutrinos we consider are relativistic during BBN, thus
 $(\rho_{\nu_s}/\rho_{\nu_\alpha})$ is the ratio during BBN of the relativistic energy densities of the sterile neutrino, $\rho_{\nu_s} = \langle\epsilon\rangle T n_{\nu_s}$ 
 with $ \langle\epsilon\rangle$ given in Eq.~\eqref{eq:nonreseps},  and one active neutrino species $\nu_{\alpha}$, $\rho_{\nu_\alpha} = 3.15 T n_{\nu_\alpha}$\footnote{We note that only for neutrinos much heavier than those we consider here sterile neutrinos could decay before BBN, thus increasing $N_{\rm eff}$ due to their decay products (heavy sterile neutrinos have been recently suggested as a solution to the observed tension between local and early Universe measurements of the Hubble constant~\cite{Gelmini:2019deq}).}. Hence,
\begin{equation} 
\label{eq:neffeq}
\Delta N_{\rm eff} = N_{\rm eff} - 3.045 \simeq  \Big(\dfrac{\langle \epsilon \rangle}{3.15}\Big) \Big(\frac{10.75}{g_\ast}\Big)^{1/3}\Big(\dfrac{n_{\nu_s}}{n_{\nu_{\alpha}}}\Big)~.
\end{equation}
Here, the ratio ${n_{\nu_s}}/n_{\nu_{\alpha}}$ in each cosmology (see Sec.~\ref{ssec:numberdensity} and  Appendix~\ref{ssecapp:numden} for sterile neutrinos produced non-resonantly) is the same during BBN and at present, since both  the number densities of sterile and of active neutrinos  just redshift for $T <$ 1 MeV (and BBN starts at about $T = 0.8$ MeV).  Using Eq. \eqref{eq:activenumden},  the present number density of one active neutrino is
\begin{equation}
\label{eq:activenumden2}
n_{\nu_{\alpha}}
\simeq 112~\textrm{cm}^{-3}~ \left(\frac{T_{\nu, 0}}{1.95~{\rm K}}\right)^3~.
\end{equation}

 The BBN upper bound of $N_{\rm eff} < 3.4$ at 95\% confidence level~\cite{Tanabashi:2018oca} through the Eq.~\eqref{eq:neffeq} rejects  the cyan region in mass-mixing labelled ``BBN" in Figs.~\ref{fig:allDWlim} and     \ref{fig:lrDWlim}.  This upper limit is similar to the Planck 2018 limit~\cite{Aghanim:2018eyx} derived from cosmic microwave background radiation (CMB) data on $N_{\rm eff}$. However, the CMB limit applies only to neutrinos that are relativistic during recombination, i.e. whose mass is $m_s \ll 1$ eV. Thus, the limit would apply only to a small region of the large parameter space we consider and hence we do not include it. In addition,  the limits imposed on the effective sterile mass $m_{s, {\rm eff}}$ or the sum of active neutrino masses~\cite{Aghanim:2018eyx, Choudhury:2018sbz} by Planck 2018, BICEP2/Keck and BAO data  do not significantly change those we obtained based on $N_{\rm eff}$ during BBN and thus we do not include them either.  Earlier Planck limits on both $N_{\rm eff}$ and $m_{s, {\rm eff}}$ were considered in a previous related study of Ref.~\cite{Rehagen:2014vna}.  

Alternatively to obtaining the limit from Eq.~\eqref{eq:neffeq}, we can use the same equation to derive a limit on the present fraction of the DM consisting of sterile neutrinos. The present value of the number density  $n_{\nu_s0}$ determines the present relic density $\rho_{s,0} =  n_{\nu_s0} m_s$  and thus the present fraction of the DM in sterile neutrinos corresponding to a particular $\Delta N_\textrm{eff}$:
\begin{equation}
\label{eq:DWNeff}
    f_{s, {\rm DM}}=
    \frac{n_{\nu_s0} m_s}{\rho_{\rm DM}} 
    =35.5 \left(\frac{\left<\epsilon\right>}{3.15}\right)^{-1} \left(\frac{m_s}{\textrm{keV}}\right)\left(\frac{T_{\nu,0}}{1.95~{\rm K}}\right)^3
    \left(\frac{g_\ast}{10.75}\right)^{\frac{1}{3}}\left(\frac{\Omega_\textrm{DM}h^2}{0.1198}\right)^{-1}  \Big(\dfrac{\Delta N_\textrm{eff}}{0.4}\Big) ~,
\end{equation} 
where $g_\ast$ is the number of entropy degrees of freedom when $\nu_s$ are produced. Using Eq.~\eqref{eq:genrhonus} for $f_{s, {\rm DM}}$,  or the equation corresponding to a particular cosmological model, $\Delta N_\textrm{eff}<0.4$ imposes an upper limit on the fraction of the DM consisting of sterile neutrinos.

\subsection{Distortions of the CMB spectrum}
\label{ssec:CMBbounds}

Photons produced in the decays of sterile neutrinos before recombination, i.e.~$\tau< t_{\rm rec}\simeq 1.2 \times 10^{13}$ sec, can  distort the CMB spectrum~\cite{Ellis:1990nb, Hu:1993gc} if they are produced after the thermalization time $t_{\rm th} \simeq 10^6$ sec (see e.g. the discussion in Ref.~\cite{Gelmini:2008fq}). 
Lifetimes of Majorana sterile neutrinos\footnote{For Dirac neutrinos the lifetime must be multiplied by 2.} equal to the lifetime of the Universe, $t_U= 4.36 \times 10^{17}$ sec, to $t_{\rm rec}$ and to $t_{\rm th}$ are indicated with red long-dashed straight lines in Figs.~\ref{fig:allDWlim} and~\ref{fig:lrDWlim}. 
The COBE FIRAS limits~\cite{Fixsen:1996nj} on distortions of the CMB spectrum reject lifetimes $t_{\rm rec}> \tau > t_{\rm th}$ (region diagonally hatched in red in Figs.~\ref{fig:allDWlim} and \ref{fig:lrDWlim} labeled ``CMB"). 

Non-thermal photons produced before the thermalization time $t_{\rm th}$ are  rapidly incorporated into the Planck spectrum through processes that change the number of photons, such as double Compton scattering ($\gamma e \to \gamma \gamma e$), which are no longer effective after $t_{\rm th}$. For $t_{\rm th} < \tau_s < 10^9~{\rm sec}$, photon number
preserving processes, such as elastic Compton scattering, are still  efficient. These processes thermalize the photons into a Bose-Einstein spectrum with a non-zero photon chemical  potential $\mu$. If the initial spectrum has fewer photons than a  black body of the same total energy the chemical potential is positive and $\mu>0$ (if instead it has more photons, $\mu<0$). For $|\mu| \ll 1$, the only
values of $\mu$ allowed by the COBE satellite limit $|\mu| < 0.9 \times
10^{-4}$ at the 95\% CL~\cite{Fixsen:1996nj}, the
energy released into photons in the decay  is~\cite{Ellis:1990nb} 
\begin{equation}
\frac{\Delta\rho_\gamma}{\rho_\gamma} \simeq  0.714 \mu~.
\label{Deltarho1}
\end{equation}
 For longer lifetimes, the photon number preserving processes can no longer establish a Bose-Einstein spectrum. Thus, for  $10^9~{\rm sec} < \tau_s < t_{\rm rec}$ the energy released into photons by the decays is not thermalized but still heats up the electrons. Through inverse-Compton scattering this
produces a distorted spectrum characterized by a  parameter $y$. The COBE bound on this parameter is $|y| < 1.5 \times 10^{-5}$~\cite{Fixsen:1996nj} and, for  $|y| \ll 1$, $y$ is related to the energy released in non-thermal photons as~\cite{Ellis:1990nb}  
\begin{equation}
\frac{\Delta\rho_\gamma}{\rho_\gamma} \simeq 4y~.
\label{Deltarho2}
\end{equation}
Coincidentally, the upper limits on $\mu$ and $y$ are such that in both cases, Eqs.~(\ref{Deltarho1}) and (\ref{Deltarho2}), the upper limit on the fractional increase in the photon energy
density due to the decay of the sterile neutrinos is $6 \times 10^{-5}$.  Thus, assuming that the decays happen instantaneously at $t=\tau_s$, and noting that the energy of each photon
produced in a decay is $m_s/2$, $\rho_\gamma = 2.7~ T~ n_\gamma$, the temperature-time relation is $T\simeq {\rm MeV}~({t}/{\rm  sec})^{1/2}$, we have
\begin{equation}
\frac{\Delta\rho_\gamma}{\rho_\gamma} =  
\frac{ B\, \,(m_s/2) \, n_{\nu_s}}{2.7~ T\, n_\gamma} \, \simeq  
\frac{ (B/2)\,m_s \, n_{\nu_s}}{2.7 \,{\rm MeV}~ n_\gamma} \, \left(\frac{\tau_s}{\rm
  sec}\right)^{1/2} \simeq  
\frac{(B/2)\, f_{s, {\rm DM}} ~ \rho_{\rm DM}}{2.7 \,{\rm MeV}~ n_\gamma} \, \left(\frac{\tau_s}{\rm
  sec}\right)^{1/2} \lesssim 6 \times 10^{-5}~. 
\end{equation}
Here  $B$ is the branching ratio for the radiative decay, so that $B~ n_{\nu_s}$ is the number of photons produced when the sterile neutrinos decay, the ratio $(n_{\nu_s}/ n_\gamma)$ is the same for any $T<1$ MeV, and  at present the energy density in sterile neutrinos is $n_{\nu_s} m_s=f_{s, {\rm DM}} ~ \rho_{\rm DM}$. In our case the branching ratio is $B=0.78 \times 10^{-2}$,  and at present $n_\gamma = 413$/cm$^3$  and $\rho_{\rm DM}\simeq 1.25$ keV/cm$^3$, thus we get
\begin{equation}
f_{s, {\rm DM}} \, \left(\frac{\tau_s}{t_{\rm rec}}\right)^{1/2} \lesssim  4 \times 10^{-3}~. 
\end{equation}
This mean that for $\tau= 10^{6}$  the limit is 
$f_{s, {\rm DM}} < 12$.
We can easily see in Figs.~\ref{fig:allDWlim} and \ref{fig:lrDWlim} that  the values of the fraction $f_{s, {\rm DM}}$ of the DM in sterile neutrinos are much larger than the upper limit in the whole lifetime range $t_{\rm rec}> \tau > t_{\rm th}$ where the limit applies.

\subsection{SN1987A disfavored region}
\label{ssec:SNbounds}
 
 The energy loss due to sterile neutrinos produced in core collapse supernovae explosions disfavors~\cite{Kainulainen:1990bn} the region
 horizontally hatched in brown and labeled ``SN"  in Figs.~\ref{fig:allDWlim} and~\ref{fig:lrDWlim}.
 Due to the considerable uncertainty in the neutrino transport and flavor transformation within hot and dense nuclear matter~\cite{Abazajian:2001nj}, it is difficult to exclude this region entirely. Recent studies regarding sterile neutrinos mixing with  $\nu_\mu$ or $\nu_\tau$ have been carried out in e.g.~Ref.~\cite{Raffelt:2011nc,Arguelles:2016uwb}.

\subsection{X-ray observations and the 3.5 keV line}
\label{ssec:xraybounds}

The most restrictive limits on sterile neutrinos with mass $m_s= 1 - 10$ keV come from astrophysical indirect detection searches of X-rays produced in their $\nu_s \rightarrow \nu_\alpha \gamma$ two-body decay\footnote{The branching ratio for the photon decay channel is subdominant, the dominant decay channel is $\nu_s \rightarrow 3\nu_\alpha$.}~\cite{Ng:2019gch,Perez:2016tcq,Neronov:2016wdd}. The rate of  this decay mode is~\cite{Shrock:1974nd, Pal:1981rm}
\begin{equation}
\label{eq:xraydecay}
    \Gamma_\gamma = 1.38\times10^{-32} ~\text{s}^{-1} \left(\frac{\sin^2 (2\theta)}{10^{-10}}\right)\left(\frac{m_s}{\textrm{keV}}\right)^{5}~.
\end{equation}
Due to the rapid decrease of the decay rate with  decreasing $m_s$, X-ray observations do not provide meaningful constraints for sterile neutrinos with $m_s < 1$ keV.

 The  model independent X-ray bounds found in the literature~\cite{Ng:2019gch,Perez:2016tcq,Neronov:2016wdd}
assume that sterile neutrinos constitute the entirety of the DM. In order to translate the published limits into those that apply in our scenarios we take into account that the X-ray signal depends on the produced photon flux, thus the limits actually constrain the product $ (\Omega_s/ \Omega_{\rm DM}) \sin^22\theta$ and not just $\sin^2 2\theta$. The present sterile neutrino fraction  $f_{s,\textrm{DM}}= (\Omega_s/ \Omega_{\rm DM})$ of the DM is itself proportional to $\sin^2 2\theta$ (because of the dependence of ${n_{\nu_s}}$ in Eq.~\eqref{eq:numden}, and Sec.~\ref{ssec:numberdensity} and  Appendix~\ref{ssecapp:numden} on the mixing).  Hence, the X-ray limits shown in Fig.~\ref{fig:allDWlim} are related to the published model independent X-ray bounds  through a simple  rescaling.

 From Eq.~\eqref{eq:overden}, DM fraction in sterile neutrinos can be written for non-resonant production as $f_{s,\textrm{DM}}=$ $\sin^2(2\theta)/ \sin^2(2\theta)_\textrm{DW,lim}$, thus the constraint on our models is the geometric mean of the published model independent limit and $\sin^2(2\theta)_\textrm{DW,lim}$ (given in Eq.~\eqref{eq:Stdoverdenselimit}, Eq.~\eqref{eq:st1overdenselimit} and Appendix~\ref{ssecapp:overdensity} for the different cosmologies we consider). 

 The same rescaling applies to upper limits on the diffuse extragalactic background radiation (DEBRA)~\cite{Boyarsky:2005us} on sterile neutrinos decaying after recombination, i.e. $\tau> t_{\rm rec}$ (the lower boundary of the DEBRA rejected region is indicated by a straight green line in Figs.~\ref{fig:allDWlim} and \eqref{fig:lrDWlim}). X-ray limits coming from galaxies and galaxy clusters only apply to relatively recent times after these structures formed, thus they are superseded by the DEBRA limits which apply on the integrated flux of all decays that occur between $t_{\rm rec}$ and the present.  
 
The $3.5$ keV X-ray emission line signal reported in 2014~\cite{Bulbul:2014sua,Boyarsky:2014jta}, which remains a matter of lively debate~(see e.g.~\cite{Dessert:2018qih,Boyarsky:2018ktr}),  could be due to the decay of $m_s\simeq$ 7 keV sterile neutrinos whose mixing should be $\sin^22\theta=5\times10^{-11}$ if they constitute all of DM. The active-sterile mixing  necessary  for sterile neutrinos  produced non-resonantly to reproduce the putative signal line shifts in the same way as the X-ray bounds just mentioned, and  is indicated with a black star in Figs.~\eqref{fig:allDWlim} and \eqref{fig:lrDWlim}. These figures show that in all the models we consider, except ST2,  the signal would correspond to sterile neutrinos that constitute only a small fraction of the DM.   Fig.~\ref{fig:allDWlim} shows that in the ST2 model, the signal rejected by Lyman-$\alpha$ limits.  Comparing  the K, ST1 and LRT to the standard cosmology in Figs.~\ref{fig:allDWlim} and \eqref{fig:lrDWlim}, we see clearly that decreased production in  cosmologies  with faster expansion rates increases the mixing angle required to produce the signal. In the K cosmology the mixings necessary to produce the 3.5 keV signal is $\sin^2 2 \theta = \mathcal{O}(10^{-9})$, and  in the LRT and ST1 cosmologies it is  $\sin^2 2 \theta = \mathcal{O}(10^{-7})$,   within the  reach of the KATRIN experiment with its proposed TRISTAN upgrade~\cite{Mertens:2018vuu} as well as   the upcoming HUNTER experiment and its upgrades~\cite{Smith:2016vku}.

\subsection{Laboratory experiments}

For eV- and keV-mass sterile neutrinos, multiple 
laboratory experiments can probe and restrict sizable portions of the parameter space. Since these experiments directly probe the active-sterile mixing angle and mass by searching for active neutrino appearance and disappearance, the resulting bounds they set are independent of cosmology and require no further modification.

In the eV-mass range of sterile neutrinos mixing with $\nu_e$, the limits are dominated by the Daya Bay~\cite{An:2016luf}, Bugey-3~\cite{Declais:1994su} and PROSPECT~\cite{Ashenfelter:2018iov} reactor and accelerator experiments, which combined reject the green regions labelled
``R" in Figs.~\ref{fig:allDWlim} and \eqref{fig:lrDWlim}. Several anomalous results, consistent with a sterile neutrino of $m_s \sim $ eV mass contributing to active-sterile neutrino oscillations, have been reported from the short-baseline studies of $\nu_e$ appearance by the LSND~\cite{Aguilar:2001ty} and MiniBooNE~\cite{Aguilar-Arevalo:2013pmq} experiments. These claims have been further bolstered recently with analysis of additional  MiniBooNE data~\cite{Aguilar-Arevalo:2018gpe}.
We show in Figs.~\ref{fig:allDWlim} and \eqref{fig:lrDWlim} with a black densely hatched band denoted ``MB" the parameter space allowed at the 4-$\sigma$ level consistent with these excesses, reproduced from  Fig.~4 of Ref.~\cite{Aguilar-Arevalo:2018gpe}  (see discussions in Ref.~\cite{Aguilar-Arevalo:2018gpe} for details). We stress, however, that the anomalous $\nu_e$ appearance results discussed above are in strong tension with the $\nu_{\mu}$ disappearance results from IceCube~\cite{TheIceCube:2016oqi} and MINOS~\cite{Adamson:2017uda}. In the same sterile neutrino mass region, recent reactor neutrino results from the DANSS~\cite{Alekseev:2018efk} and NEOS~\cite{Ko:2016owz} experiments are also consistent with a sterile neutrino interpretation. A combined fit to their data~\cite{Dentler:2018sju,Gariazzo:2018mwd,Liao:2018mbg} allows for a sterile neutrino with $m_s \simeq 1.14$ eV mass and  $\sin^2 2\theta \simeq 0.04$ mixing with a $\nu_e$ active neutrino. The regions allowed at 3-$\sigma$ level by the DANSS and NEOS data reproduced from Fig. 4 of Ref.~\cite{Gariazzo:2018mwd} are indicated with black vertical elliptical  contours in Figs.~\ref{fig:allDWlim} and \ref{fig:lrDWlim}.

 As we display in Fig.~\ref{fig:allDWlim} and \eqref{fig:lrDWlim}, the eV-mass parameter space relevant for anomalous observations in short-baseline and reactor experiments will be fully or at least partially tested by KATRIN~\cite{megas:thesis} (whose reach is shown with a solid blue line labeled ``KA")  as well as PTOLEMY~\cite{Betti:2019ouf} (whose reach is shown with solid orange lines labeled ``P" and ``P2" for 10 mg-yr and 100 g-yr exposures, respectively). PTOLEMY is a tritium $\beta$-decay experiment aimed at detecting the cosmological relic neutrino background  which is expected to start collecting data within few years. As shown in Fig.~\ref{fig:allDWlim} and Fig.~\ref{fig:lrDWlim}, the cosmological bounds in the ST1 and LRT cosmologies are significantly relaxed for eV-mass sterile neutrinos compared to those in the Std cosmology. Hence, the required parameter space for anomalous observations in short-baseline and reactor experiments is not rejected by cosmology.

In the keV mass-scale, the tritium decay experiment KATRIN~\cite{Wolf:2008hf,Mertens:2015ila} will probe active-sterile mixing down to $\sin^2(2\theta)\leq 10^{-4}$~\cite{Mertens:2018vuu}, shown with solid blue and denoted ``KA'' on Figs.~\ref{fig:allDWlim} and \ref{fig:lrDWlim}. Its upgraded version, TRISTAN (denoted with a solid blue line and labelled ``T'' and ``T2'', corresponding to a 1 year and 3 year data collecting period, respectively), is expected to reach sensitivities of $\sin^2 (2 \theta) \sim  10^{-8}$ within a 3 year run-time~\cite{Mertens:2018vuu}. The upcoming cesium trap experiment HUNTER~\cite{Smith:2016vku} is expected to probe even more further within the sterile neutrino parameter space. Here, the missing mass of the neutrino will be reconstructed from $^{131}$Cs electron capture  decays occurring in a magneto-optically trapped sample.
The prototype version of the experiment is already under construction and will have two phases, whose sensitivity we display in Figs.~\ref{fig:allDWlim} and \eqref{fig:lrDWlim} with magenta solid lines and labelled ``H1A'' and ``H1B'', respectively. The upgrade version of Hunter, denoted by ``HU'', is expected to be sensitive to mixings down to 
 $\sin^2 (2 \theta) \sim 10^{-11}$. The above-mentioned experiments will be able to test the sterile neutrino origin of the 3.5 keV X-ray signal line. In non-standard cosmologies that result in decreased sterile neutrino density, the required mixing angle to explain the signal increases, allowing for sterile neutrinos to appear more visible for laboratory studies (see e.g. LRT and ST1 in Figs.~\ref{fig:allDWlim} and \eqref{fig:lrDWlim}).
 
 If the sterile neutrinos are Majorana particles, they will mediate the neutrinoless double beta $0\nu\beta\beta$ decay.
 A sterile neutrino that mixes with the electron neutrino contributes   $\left<m\right>_s = m_s \sin^2(\theta) e^{i\beta_s}$ to the effective electron neutrino Majorana mass $\left<m\right>$  that affects the half-life of $0\nu\beta\beta$ decay, where $\beta_s$ denotes a Majorana CP-violating phase. Hence, using the present bound on the magnitude of  $\left<m\right>$, $|\langle m\rangle|<0.165$ eV~\cite{KamLAND-Zen:2016pfg}, the corresponding bound on the sterile neutrino mass and mixing angle is $m_s\sin^2(2\theta)<0.660$ eV. This limit is shown in orange in Figs.~\ref{fig:allDWlim} and \eqref{fig:lrDWlim}, with the label ``$0\nu\beta\beta$".  We note that this bound is not completely robust. The contribution of the sterile neutrino might interfere with the contributions from the active ones, leading to a suppression in the effective Majorana mass and, therefore, avoiding the experimental bounds~\cite{Abada:2018qok}.   
 
\section{Summary}
\label{sec:summary}

The early Universe pre-BBN cosmology could be drastically different from the usually assumed radiation-dominated cosmology with SM particle content, as happens in  motivated theoretical models -- for example in models based on moduli or quintessence. Since no remnant has been detected from the pre-BBN epoch, this era of the early Universe currently remains completely untested. Visible sterile neutrinos, those which could be detected in near future laboratory experiments, could be the first remnants from this epoch.  Here we revisited the  production of sterile neutrinos via  non-resonant active-sterile oscillations assuming different cosmologies before the temperature of the Universe was 5 MeV and showed that these neutrinos can act as sensitive probes of the pre-BBN cosmology.

In particular,  we studied non-resonant sterile neutrino production within the standard and several non-standard cosmologies before $T=5$ MeV. We dealt mostly with cosmological models in which entropy in matter and radiation is conserved, such as the Scalar Tensor and Kination models, using a parameterization of the expansion rate $H$ in terms of its amplitude and temperature dependence that has the particular cosmologies we studied as special cases. We also revisited Low Reheating Temperatures models, in which entropy is not conserved during the non-standard cosmological phase, but sterile neutrinos are produced dominantly in the standard phase, at $T< 5$ MeV. In all cases we assumed that the cosmology is standard at $T< 5$ MeV.  

We found that the resulting sterile neutrino relic abundance can be either suppressed or enhanced  and that the momentum distribution can be colder or hotter compared to those in the standard cosmology for the same neutrino masses and mixings. We derived general expressions for all relevant quantitites  using the mentioned  parametrization of $H$ with two parameters, and give them also for the particular cosmological models we studied. We updated and extended the cosmological and astrophysical bounds on all the cosmologies we considered. 

In particular, in the Low Reheating Temperature  and one of the Scalar Tensor (ST1) models  we studied, the cosmological bounds are significantly relaxed, and the mixing of the sterile neutrinos possibly responsible for the 3.5 keV signal is more accessible to the reach of the upcoming KATRIN/TRISTAN and  HUNTER experiments. These experiments have already started or are expected to start taking data soon. The observation of a $\sim 7$ keV mass sterile neutrino in one of them would not only constitute a momentous discovery in particle physics, but also in cosmology, even if this neutrino does not consitute all of the dark matter. Namely, it would not only constitute the discovery of a new elementary particle,  a particle physics discovery of fundamental importance, but could hold vital information about the pre-BBN cosmology from which this sterile neutrino could be the first ever detected remnant.

For example, if the measured mixing would be $\sin^2 2 \theta = \mathcal{O}(10^{-7})$, the discovery would be consistent with a  non-standard cosmology such as the ST1 and Low Reheating Temperature we studied here. On the other hand, a measured value of  $\sin^2 2 \theta = \mathcal{O}(10^{-9})$ could instead point towards a Kination model (or maybe special particle models, e.g.~\cite{Bezrukov:2017ike}).
 For a $\sin^2 2 \theta = \mathcal{O}(10^{-10})$, it would point to a  standard pre-BBN cosmology, as indicated in Figs.~\ref{fig:allDWlim} and \eqref{fig:lrDWlim} (see the location of the black stars).

The relaxation of cosmological bounds on eV-scale mass sterile neutrinos in non-standard pre-BBN cosmologies, mostly in the Low Reheating Temperature and the ST1 models, allows for the results reported from the LSND, and MiniBooNE short-baseline as well as the DANSS and NEOS reactor neutrino experiments to be unrestricted by cosmology\footnote{The cosmological limits can be also suppressed by additional sterile neutrino interactions, such as those due to their coupling to an ultra-light scalar (see e.g. Refs.~\cite{Farzan:2019yvo,Cline:2019seo}).}.  If the discovery of a sterile neutrino in any of these experiments would be confirmed, again this would be of fundamental importance not only for particle physics, but possibly for the pre-BBN cosmology in which they were produced, and it could provide an indication of a non-standard cosmology in this yet untested epoch.

\acknowledgments
\addcontentsline{toc}{section}{Acknowledgments}
 
The work of G.B.G., P.L. and V.T. was supported in part by the U.S. Department of Energy (DOE) Grant No. DE-SC0009937.  

\appendix
\section{Additional formulas for non-resonant production}
\label{sec:appendixformulas}

Below we show the equations in their general form as function of the $\eta$ and $\beta$ parameters appearing in the  parametrization of $H$ given in Eq.~\eqref{eq:hnStd} and for the K,  ST2 and LRT cosmologies, which were not given in the main text. 

We give the equations for the following quantities: the temperature of maximum rate of production $T_{\textrm{max}}$ (except for  the LRT model in which the maximum rate of production is at $T_{\rm RH}$), the sterile neutrino momentum distributions $f_{\nu_s}(\epsilon)$ as function of the diensional momentum $\epsilon = p/T$, the relic number density $n_{\nu_s}$ and the  relic energy density $\rho_{\nu_s}$ in general and at present, the present fraction $f_{s, {\rm DM}}$ of the DM consisting of sterile neutrinos, and the mixing and function of mass $\sin^2(2\theta)_\textrm{DW,lim}$ for which this fraction is 1. 

In the main text we included only the Std and the ST1 to easily compare the standard cosmology with the alternative cosmology that provides the largest departure from the standard  results. 
 
\subsection{Temperature of maximum rate of production of sterile neutrinos}
\label{ssecapp:maxtemp}
 
For the parametrization of $H$ in Eq.~\eqref{eq:hnStd} the temperature of maximum DW production for $\beta \leq 2$ is
\begin{align}
\begin{split}\label{eq:mcmax}
    T_{\textrm{max}}   
    \simeq 190 \textrm{ MeV}~\epsilon^{-\frac{1}{3}} \left(\frac{2-\beta }{10+\beta }\right)^\frac{1}{6} \left(\frac{m_s}{\textrm{keV}}\right)^\frac{1}{3} \left(\frac{B}{10.88\times10^{-9}\textrm{ GeV}^{-4}}\right)^{-\frac{1}{6}}~.
\end{split}
\end{align}
In K pre-BBN cosmology ($\beta = 1$) it is
\begin{equation} \label{eq:ktmax}
    T_{\textrm{max}}^{\rm K} = 127 \textrm{ MeV}~ \epsilon^{-\frac{1}{3}}\left(\frac{m_s}{\textrm{keV}}\right)^{\frac{1}{3}} \left(\frac{B}{10.88\times10^{-9}\textrm{ GeV}^{-4}}\right)^{-\frac{1}{6}}~,
\end{equation}
and the ST2 cosmology ($\beta = 0$),
\begin{equation} \label{eq:st2tmax}
    T_{\textrm{max}}^{\rm ST2} = T_{\rm max}^{\rm Std} = 145 \textrm{ MeV}~\epsilon^{-\frac{1}{3}} \left(\frac{m_s}{\textrm{keV}}\right)^{\frac{1}{3}} \left(\frac{B}{10.88\times10^{-9}\textrm{ GeV}^{-4}}\right)^{-\frac{1}{6}}~.
\end{equation}
The $T_{\rm max}$ for ST2 and the standard cosmology coincide, because $\beta = 0$ for both. 

\subsection{Momentum distribution functions of non-resonantly produced sterile neutrinos}
\label{ssecapp:distfunc}
For the parametrization of $H$ in Eq.~\eqref{eq:hnStd} the momentum distribution function, as function of $\epsilon = p/T$, is given by
\begin{align}
\label{eq:mcdist2}
    f_{\nu_s}(\epsilon) ~=&~ 3.08 \times 10^{-6} \,\eta^{-1} \epsilon^{\frac{\beta}{3}} (3+\beta) (2.63\times10^{-2})^\beta\sec\left(\frac{\beta\pi}{6}\right) \left(\frac{\sin^2(2\theta)}{10^{-10}}\right)  \notag\\ 
    &\times  \left(\frac{m_s}{\textrm{keV}}\right)^{1-\frac{\beta}{3}}\left(\frac{T_{\rm tr}}{5 \textrm{ MeV}}\right)^\beta  \left(\frac{g_{\ast}}{30}\right)^{-\frac{1}{2}} \left(\frac{d_\alpha}{1.13}\right) \notag\\
    &\times \left(\frac{B}{10.88\times10^{-9}\textrm{ GeV}^{-4}}\right)^{-\frac{1}{2}+\frac{\beta}{6}}f_{\nu_\alpha}(\epsilon)~.
\end{align}
In the K cosmology $\eta = 1$, $\beta = 1$, thus
\begin{align}
\label{eq:kdist}
    f_{\nu_s}^{\rm K}(\epsilon) ~=&~ 3.74\times10^{-7} \epsilon^{\frac{1}{3}}\left(\frac{\sin^2(2\theta)}{10^{-10}}\right)\left(\frac{m_s}{\textrm{keV}}\right)^{\frac{2}{3}} \left(\frac{T_{\rm tr}}{5 \textrm{ MeV}}\right)\left(\frac{g_{\ast}}{30}\right)^{-\frac{1}{2}} \notag\\
    &\times\left(\frac{d_\alpha}{1.13}\right) \left(\frac{B}{10.88\times10^{-9}\textrm{ GeV}^{-4}}\right)^{-\frac{1}{3}}f_{\nu_\alpha}(\epsilon) 
\end{align}
and in the ST2 cosmology, $\eta = 0.03$ and $\beta = 0$, instead
\begin{align}
\label{eq:st2dist}
    f_{\nu_s}^{\rm ST2}(\epsilon) ~=&~ 2.89\times10^{-4}\left(\frac{\sin^2(2\theta)}{10^{-10}}\right)\left(\frac{m_s}{\textrm{keV}}\right)\left(\frac{g_{\ast}}{30}\right)^{-\frac{1}{2}} \notag\\ 
    &\times\left(\frac{d_\alpha}{1.13}\right) \left(\frac{B}{10.88\times10^{-9}\textrm{ GeV}^{-4}}\right)^{-\frac{1}{2}}f_{\nu_\alpha}(\epsilon)~.
\end{align}

The distribution function of sterile neutrinos for low reheating temperature  (LRT) cosmologies is given in Eq. (1) of Ref. \cite{Gelmini:2004ah}, for $T_{\rm RH}= T_\textrm{tr}$. In these models sterile neutrinos are dominantly produced during the radiation-dominated period, i.e. for $T < T_{\rm RH}$. We reproduce it here for completeness,
\begin{align}
\label{eq:lrtdist}
    f_{\nu_s}^{\textrm{LRT}} = 3.6\times10^{-10} \,  \epsilon \, \left(\frac{\sin^2(2\theta)}{10^{-10}}\right)\left(\frac{T_\textrm{tr}}{5\textrm{ MeV}}\right)^3 \left(\frac{d_\alpha}{1.13}\right)f_{\nu_\alpha}(\epsilon)~.
\end{align}
For consistency of notation, here use $T_{\rm tr}$ for the reheating temperature.

\subsection{Relic number density of non-resonantly produced sterile neutrinos}
\label{ssecapp:numden}

The ratio of the number density of non-resonantly produced sterile neutrinos and of active neutrinos at the same temperature $T$ is easily related to the ratio of their momentum distributions,  
\begin{align}\label{eq:numratio-gen}
    \frac{n_{\nu_s}(T)}{n_{\nu_a}(T)} = \epsilon^{- \frac{\beta}{3}}\, 
    \frac{F_{2+\frac{\beta}{3}}(0)}{F_{2}(0)}\frac{f_{\nu_s}(\epsilon)}{f_{\nu_\alpha}(\epsilon)}.
\end{align}
where we have used the  $\eta$ and $\beta$ parametrization of $H$ and the number density  of active neutrinos (and antineutrinos) here is $n_{\nu_{\alpha}} = \left({3\zeta(3)}/{2\pi^2}\right)T^3$. For each cosmology, this ratio is
\begin{equation}
\label{eq:numratio}
 \frac{n_{\nu_s}(T)}{n_{\nu_\alpha}(T)}
= \frac{f_{\nu_s}(\epsilon)}{f_{\nu_\alpha}(\epsilon)} \times \left \{
  \begin{tabular}{rl}
  1, & ~~~\text{Std, ST2}\\
  1.42~$\epsilon^{-0.33}$, & ~~~\text{K} \\
  0.77~$\epsilon^{~0.27}$~, & ~~~\text{ST1} \\
  3.15~$\epsilon^{-1}$~~~, & ~~~\text{LRH} 
  \end{tabular}
\right .\
\end{equation}

The present number density of non-resonantly produced sterile neutrinos as a function of the $\eta$ and $\beta$ parameters of $H$ in Eq.~\eqref{eq:hnStd} is
\begin{align}\label{eq:numden}
    n_{\nu_s} =&~ 7.81\times10^{-3} \textrm{cm}^{-3}\eta^{-1}\left(2.63\times10^{-2}\right)^\beta  \left(1-2^{-2-\frac{\beta }{3}}\right)\zeta\left(3+\frac{\beta }{3}\right)\Gamma\left(3+\frac{\beta }{3}\right) \notag\\ &\times \left(\frac{3+\beta }{36\pi}\right)\sec\left(\frac{\beta \pi}{6}\right)\left(\frac{\sin^2(2\theta)}{10^{-10}}\right)\left(\frac{m_s}{\textrm{keV}}\right)^{1-\frac{\beta }{3}}\left(\frac{T_{\rm tr}}{5\textrm{ MeV}}\right)^\beta\left(\frac{T_{\nu,0} }{1.95~{\rm K}}\right)^{3}   \notag\\ &\times\left(\frac{g_{\ast}}{30}\right)^{-\frac{3}{2}}\left(\frac{d_\alpha}{1.13}\right)  \left(\frac{B}{10.88\times10^{-9}\textrm{ GeV}^{-4}}\right)^{-\frac{1}{2}+\frac{\beta }{6}}~.
\end{align}
Here $g_{\ast}$ is the number of relativistic degrees of freedom when sterile neutrinos are produced, which we take to be $g_{\ast}( T_{\rm max})$ thus at present the ratio of temperatures of the sterile and active neutrinos is $(T_{\nu_s,0}/T_{\nu, 0}) = (g_{\ast}/10.75)^{1/3}$. $T_{\nu, 0} = 1.95$ K is the present temperature of the relic active neutrino bath.

\noindent For K ($\eta = 1$, $\beta = 1$) the sterile neutrino number density is
\begin{align}
\label{eq:knumden-general}
    n_{\nu_s}^{\rm K}(T_{\nu_s}) ~=&~ 5.29\times10^{-7}\left(\frac{\sin^2 (2\theta)}{10^{-10}}\right)\left(\frac{m_s}{\textrm{keV}}\right)^{\frac{2}{3}}\left(\frac{T_{\rm tr}}{5\textrm{ MeV}}\right)\left(\frac{T_{\nu_s}}{T_{\nu_\alpha}}\right)^3\left(\frac{g_{\ast}}{30}\right)^{-\frac{1}{2}} \notag\\
    &\times\left(\frac{d_\alpha}{1.13}\right)  \left(\frac{B}{10.88\times10^{-9}\textrm{ GeV}^{-4}}\right)^{-\frac{1}{3}} n_{\nu_\alpha}(T_{\nu_\alpha})~,
\end{align}
and the present number density is
\begin{align}
\label{eq:knumden}
    n_{\nu_s}^{\rm K}=&~ 2.14\times10^{-5}\textrm{cm}^{-3}\left(\frac{\sin^2 (2\theta)}{10^{-10}}\right)\left(\frac{m_s}{\textrm{keV}}\right)^{\frac{2}{3}}\left(\frac{T_{\rm tr}}{5\textrm{ MeV}}\right)\left(\frac{T_{\nu, 0}}{1.95~{\rm K}}\right)^3 \notag\\
    &\times \left(\frac{g_{\ast}}{30}\right)^{-\frac{3}{2}} \left(\frac{d_\alpha}{1.13}\right)\left(\frac{B}{10.88\times10^{-9}\textrm{ GeV}^{-4}}\right)^{-\frac{1}{3}}~.
\end{align}
For ST2 ($\eta = 0.03$, $\beta = 0$) the number density it
\begin{align}
\label{eq:st2numden-general}
    n_{\nu_s}^{\rm ST2}(T_{\nu_s}) ~=&~ 2.89\times10^{-4}\left(\frac{\sin^2 (2\theta)}{10^{-10}}\right)\left(\frac{m_s}{\textrm{keV}}\right)\left(\frac{T_{\nu_s}}{T_{\nu_\alpha}}\right)^3\left(\frac{g_{\ast}}{30}\right)^{-\frac{1}{2}} \notag\\ 
    &\times  \left(\frac{d_\alpha}{1.13}\right)\left(\frac{B}{10.88\times10^{-9}\textrm{ GeV}^{-4}}\right)^{-\frac{1}{2}} n_{\nu_\alpha}(T_{\nu_\alpha}) ~,
\end{align}
and the present number density is
\begin{align}
\label{eq:st2numden}
     n_{\nu_s}^{\rm ST2}=&~ 1.17\times10^{-2}\textrm{cm}^{-3}\left(\frac{\sin^2 (2\theta)}{10^{-10}}\right)\left(\frac{m_s}{\textrm{keV}}\right)\left(\frac{T_{\nu, 0}}{1.95~{\rm K}}\right)^3\notag\\ 
    &\times\left(\frac{g_{\ast}}{30}\right)^{-\frac{3}{2}}\left(\frac{d_\alpha}{1.13}\right)   \left(\frac{B}{10.88\times10^{-9}\textrm{ GeV}^{-4}}\right)^{-\frac{1}{2}}~.
\end{align}
In the LRT cosmology (see Eq.~2 of Ref.~\cite{Gelmini:2004ah}) the number density is
\begin{align}
\label{eq:lrtnumden-general}
    n_{\nu_s}^{\textrm{LRT}}(T_{\nu_s}) ~=~ & 1.13\times10^{-9} \left(\frac{\sin^2(2\theta)}{10^{-10}}\right)\left(\frac{T_\textrm{tr}}{5\textrm{ MeV}}\right)^3\left(\frac{d_\alpha}{1.13}\right) n_{\nu_\alpha}(T_{\nu_\alpha})~,
\end{align}
and the present number density is
\begin{align}
\label{eq:lrtnumden}
    n_{\nu_s}^{\textrm{LRT}} ~=~ & 1.28\times10^{-7} \textrm{cm}^{-3}\left(\frac{\sin^2(2\theta)}{10^{-10}}\right)\left(\frac{T_\textrm{tr}}{5\textrm{ MeV}}\right)^3\left(\frac{T_{\nu,0}}{1.95\textrm{ K}}\right)^3\left(\frac{d_\alpha}{1.13}\right)~.
\end{align}

\subsection{Energy density of non-resonantly produced relativistic sterile neutrinos}
\label{ssecapp:relenergy}

The ratio of the energy density of non-resonantly produced relativistic sterile neutrinos and of relativistic active neutrinos at the same temperature $T$ is easily related to  their number density ratio, and momentum distribution ratios,  
\begin{align}\label{eq:densratio-gen}
    \frac{\rho_{\nu_s}(T)}{\rho_{\nu_a}(T)} = \frac{F_{3+\frac{\beta}{3}}(0)}{F_{3}(0)}\frac{F_{2}(0)}{F_{2+\frac{\beta}{3}}(0)}\frac{n_{\nu_s}(T)}{n_{\nu_\alpha}(T)} =\epsilon^{- \frac{\beta}{3}}\, 
    \frac{F_{3+\frac{\beta}{3}}(0)}{F_{3}(0)}\frac{f_{\nu_s}(\epsilon)}{f_{\nu_\alpha}(\epsilon)}.
\end{align}
where the energy density  of relativistic active neutrinos (and antineutrinos) is here $\rho_{\nu_{\alpha}} =  \left(7 \pi^2/120\right)T^4$. Hence,
\begin{equation}
\label{eq:numratio}
 \frac{\rho_{\nu_s}(T)}{\rho_{\nu_\alpha}(T)}
=  \frac{n_{\nu_s}(T)}{n_{\nu_\alpha}(T)} \times \left \{
  \begin{tabular}{rl}
  1, & ~~~\text{Std, ST2}\\
  1.10, & ~~~\text{K} \\
  0.92, & ~~~\text{ST1} \\
  1.30, & ~~~\text{LRH} 
  \end{tabular}
\right .\
\end{equation}

The energy density of relativistic sterile neutrinos at temperature $T_{\nu_s}$ produced non resonantly when the number of degrees of freedom were $g^*$ (we take this to be the $g_*$ at $T_{max}$), using the  $\eta$ and $\beta$ parameterization of $H$, is
\begin{align}\label{eq:genrhonus}
    \rho_{\nu_s}(T_{\nu_s}) =&~ 4.40\times10^{26}~ \textrm{MeV}/\textrm{cm}^{3}~\eta^{-1}\left(2.63\times10^{-2}\right)^\beta  \left(1-2^{-3-\frac{\beta}{3}}\right)\zeta\left(4+\frac{\beta }{3}\right)\Gamma\left(4+\frac{\beta }{3}\right) \notag\\ &\times \left(\frac{3+\beta }{36\pi}\right)\sec\left(\frac{\beta \pi}{6}\right)\left(\frac{\sin^2(2\theta)}{10^{-10}}\right)\left(\frac{m_s}{\textrm{keV}}\right)^{1-\frac{\beta }{3}}\left(\frac{T_{\rm tr}}{5\textrm{ MeV}}\right)^\beta  \left(\frac{T_{\nu_s} }{~{\rm MeV}}\right)^{4} \notag\\ &\times\left(\frac{g_{\ast}}{30}\right)^{-\frac{11}{6}}  \left(\frac{d_\alpha}{1.13}\right)\left(\frac{B}{10.88\times10^{-9}\textrm{ GeV}^{-4}}\right)^{-\frac{1}{2}+\frac{\beta }{6}}~,
\end{align}
which for K becomes
\begin{align}
\label{eq:krhonus-1}
    \rho_{\nu_s}^{\rm K}(T_{\nu_s}) ~=&~ 5.83\times10^{-7}\left(\frac{\sin^2 (2\theta)}{10^{-10}}\right)\left(\frac{m_s}{\textrm{keV}}\right)^{\frac{2}{3}}\left(\frac{T_{\rm tr}}{5\textrm{ MeV}}\right) \left(\frac{T_{\nu_s}}{T_{\nu_\alpha}}\right)^4\left(\frac{g_{\ast}}{30}\right)^{-\frac{1}{2}}
     \notag\\
    &\times\left(\frac{d_\alpha}{1.13}\right) \left(\frac{B}{10.88\times10^{-9}\textrm{ GeV}^{-4}}\right)^{-\frac{1}{3}}~ \rho_{\nu_\alpha}(T_{\nu_\alpha})~, 
\end{align}
or, 
\begin{align}
\label{eq:krhonus-2}
    \rho_{\nu_s}^{\rm K}(T_{\nu_s})
    =&~ 4.37\times10^{25}~\textrm{MeV}/\textrm{cm}^{3}\left(\frac{\sin^2 (2\theta)}{10^{-10}}\right)\left(\frac{m_s}{\textrm{keV}}\right)^{\frac{2}{3}}\left(\frac{T_{\rm tr}}{5\textrm{ MeV}}\right)\left(\frac{T_{\nu_s}}{1~{\rm MeV}}\right)^4
    \notag\\
    &\times \left(\frac{g_{\ast}}{30}\right)^{-\frac{1}{2}} \left(\frac{d_\alpha}{1.13}\right)  \left(\frac{B}{10.88\times10^{-9}\textrm{ GeV}^{-4}}\right)^{-\frac{1}{3}}~,
\end{align}
and for ST2 is
\begin{align}
\label{eq:st2rhonus-1}
    \rho_{\nu_s}^{\rm ST2}(T_{\nu_s}) ~=&~ 2.89\times10^{-4}\left(\frac{\sin^2 (2\theta)}{10^{-10}}\right)\left(\frac{m_s}{\textrm{keV}}\right)\left(\frac{T_{\nu_s}}{T_{\nu_\alpha}}\right)^4\left(\frac{g_{\ast}}{30}\right)^{-(1/2)}\notag\\ 
    &\times \left(\frac{d_\alpha}{1.13}\right)  \left(\frac{B}{10.88\times10^{-9}\textrm{ GeV}^{-4}}\right)^{-\frac{1}{2}} \rho_{\nu_\alpha}(T_{\nu_\alpha}) ~,
\end{align}
or,
\begin{align}
\label{eq:st2rhonus-2}
    \rho_{\nu_s}^{\rm ST2}(T_{\nu_s}) 
    =&~ 2.17\times10^{28}~\textrm{MeV}/\textrm{cm}^{3}~\left(\frac{\sin^2 (2\theta)}{10^{-10}}\right)\left(\frac{m_s}{\textrm{keV}}\right)\left(\frac{T_{\nu_s}}{1~{\rm MeV}}\right)^4 \notag\\ 
    &\times \left(\frac{g_{\ast}}{30}\right)^{-(1/2)}\left(\frac{d_\alpha}{1.13}\right) \left(\frac{B}{10.88\times10^{-9}\textrm{ GeV}^{-4}}\right)^{-\frac{1}{2}}~.
\end{align}

In the LRT model, production happens only during the standard phase (i.e. when $\eta=$ 1 and $\beta=$ 0), at temperatures smaller than the reheating temperature $T_{\rm RH}$, which we denote here $T_{\rm tr}$, and is dominated by the production close to $T_{\rm tr}$.  For temperatures 
$m_s< T< T_{\rm tr}$ the relic energy density in the LRT model is
\begin{align}
\label{eq:lrtrhonus}
    \rho_{\nu_s}^{\rm LRT}(T_{\nu_s}) ~=&~ 1.48\times10^{-9}\left(\frac{\sin^2 (2\theta)}{10^{-10}}\right)\left(\frac{T_{\rm tr}}{5\textrm{ MeV}}\right)^3\left(\frac{T_{\nu_s}}{T_{\nu_\alpha}}\right)^4\left(\frac{d_\alpha}{1.13}\right)  \rho_{\nu_\alpha}(T_{\nu_\alpha}) \notag \\
    =&~ 1.11\times10^{23}~\textrm{MeV}/\textrm{cm}^{3}~\left(\frac{\sin^2 (2\theta)}{10^{-10}}\right)\left(\frac{T_{\rm tr}}{5\textrm{ MeV}}\right)^3\left(\frac{T_{\nu_s}}{1~{\rm MeV}}\right)^4\left(\frac{d_\alpha}{1.13}\right) ~.
\end{align}
 
\subsection{Present fraction of the DM in non-resonantly produced sterile neutrinos}
\label{ssecapp:abundance}

The present fraction of DM comprised of non-relativistic sterile neutrinos at present as function of the $\eta$ and $\beta$ parameters in Eq.~\eqref{eq:hnStd} is  
\begin{align}
\label{eq:overden}
    f_{s, {\rm DM}} =&~ \Big(\dfrac{n_{\nu_s,0} \, m_s}{\rho_{\textrm{DM}}}\Big) ~=~ \notag \\ =&~ 4.59\times10^{-5} \eta^{-1} \left(2.63\times10^{-2}\right)^\beta (3+\beta)\left(1-2^{-2-\frac{\beta}{3}}\right)\zeta\left(3+\frac{\beta}{3}\right)
     \notag\\ & \times\Gamma\left(3+\frac{\beta}{3}\right) \sec\left(\frac{\beta\pi}{6}\right) \left(\frac{\sin^2(2\theta)}{10^{-10}}\right)  \left(\frac{m_s}{\textrm{keV}}\right)^{2-\frac{\beta}{3}}  
    \left(\frac{T_{\rm tr}}{5\textrm{ MeV}}\right)^\beta \notag\\ & \times \left(\frac{T_{\nu,0}}{1.95~{\rm K}}\right)^3    \left(\frac{g_{\ast}}{30}\right)^{-\frac{3}{2}}\left(\frac{d_\alpha}{1.13}\right) \left(\frac{\Omega_\textrm{DM}h^2}{ 0.12}\right)
    \left(\frac{B}{10.88\times10^{-9}\textrm{ GeV}^{-4}}\right)^{-\frac{1}{2}+\frac{\beta}{6}}~, 
\end{align}
where $n_{\nu_s,0}$ is the present number density, $g_\ast$ is the number of entropy degrees of freedom when the sterile  neutrinos were produced, $g_\ast=g_\ast(T_{max})$,  and $T_{\nu,0}$ is the present active neutrino  temperature. Unless stated otherwise we take $g_\ast=30$ for our figures.  For the K cosmology ($\eta = 1$, $\beta = 1$) the fraction is
\begin{align}
\label{eq:kinoverden}
    f_{s, {\rm DM}}^{\rm K} =&~ \Big(\dfrac{n_{\nu_s,0} \, m_s}{\rho_{\textrm{DM}}}\Big)^{\rm K}  ~=~ \notag \\ =&~ 1.42\times10^{-5} \left(\frac{\sin^2(2\theta)}{10^{-10}}\right)\left(\frac{m_s}{\textrm{keV}}\right)^{\frac{5}{3}}\left(\frac{T_{\rm tr}}{5\textrm{ MeV}}\right)\left(\frac{T_{\nu,0}}{1.95~{\rm K}}\right)^3\notag\\ & \times \left(\frac{g_{\ast}}{30}\right)^{-\frac{3}{2}}\left(\frac{d_\alpha}{1.13}\right)  \left(\frac{\Omega_\textrm{DM}h^2}{ 0.12}\right)\left(\frac{B}{10.88\times10^{-9}\textrm{ GeV}^{-4}}\right)^{-\frac{1}{3}}
\end{align}
and for ST2 cosmology ($\eta = 0.03$, $\beta = 0$),
\begin{align}
\label{eq:st2overden}
     f_{s, {\rm DM}}^{\rm ST2} =&~ \Big(\dfrac{n_{\nu_s,0}\, m_s}{\rho_{\textrm{DM}}}\Big)^{\rm ST2}  ~=~ \notag \\ =&~ 7.74\times10^{-3}\left(\frac{\sin^2(2\theta)}{10^{-10}}\right)\left(\frac{m_s}{\textrm{keV}}\right)^{2}
     \left(\frac{T_{\nu, 0}}{1.95~{\rm K}}\right)^3\left(\frac{g_{\ast}}{30}\right)^{-\frac{3}{2}} \notag\\ & \times\left(\frac{d_\alpha}{1.13}\right)
        \left(\frac{\Omega_\textrm{DM}h^2}{ 0.12}\right)\left(\frac{B}{10.88\times10^{-9}\textrm{ GeV}^{-4}}\right)^{-\frac{1}{2}}~.
\end{align}
In the LRT model with $T_{\rm tr}$ denoting the reheating temperature, the fraction is instead
\begin{align}
\label{eq:lrtoverden}
    f_{s,\textrm{DM}}^{\rm LRT} =&~ \Big(\dfrac{n_{\nu_s,0} \,
    m_s}{\rho_{\textrm{DM}}}\Big)^{\rm LRT} ~=~ \notag \\ =&~ 1 \times 10^{-7}\left(\frac{\sin^2(2\theta)}{10^{-10}}\right)
    \left(\frac{m_s}{\textrm{keV}}\right)
     \left(\frac{T_{\rm tr}}{5{\rm MeV}}\right)^3 
    \left(\frac{T_{\nu,0}}{1.95~\textrm{K}}\right)^3
     \notag\\ & \times
     \left(\frac{d_\alpha}{1.13}\right)\left(\frac{\Omega_\textrm{DM} h^2}{0.12}\right) ~.
\end{align}

\subsection{DM density limit}
\label{ssecapp:overdensity}

The limit on the sterile-active neutrino mixing angle from DM density, $f_{s, {\rm DM}} \leq 1$ translates into $\sin^2 (2 \theta) < \sin^2 (2\theta)_{\rm DW, lim}$, as function of $\eta$ and $\beta$ in Eq.~\eqref{eq:hnStd}, where
\begin{align}
\label{eq:overdenselimit}
    \sin^2 (2\theta)_\textrm{DW,lim} ~=&~ 2.18\times10^{-6}\eta \left[\left(1-2^{-2-\frac{\beta}{3}}\right)\zeta\left(3+\frac{\beta}{3}\right)\sec\left(\frac{\beta\pi}{6}\right)\Gamma\left(3+\frac{\beta}{3}\right)(3+\beta)\right]^{-1} \notag\\ 
    &\times \left(2.63\times10^{-2}\right)^{-\beta}\left(\frac{m_s}{\textrm{keV}}\right)^{-2+\frac{\beta}{3}} \left(\frac{T_{\rm tr}}{5 \textrm{ MeV}}\right)^{-\beta}\left(\frac{T_{\nu,0}}{1.95 ~{\rm K}}\right)^{-3}\left(\frac{g_{\ast}}{30}\right)^{\frac{3}{2}} \notag\\ &\times \left(\frac{d_\alpha}{1.13}\right)^{-1}\left(\frac{\Omega_\textrm{DM}h^2}{ 0.12}\right)^{-1}\left(\frac{B}{10.88\times10^{-9}\textrm{ GeV}^{-4}}\right)^{\frac{1}{2}-\frac{\beta}{6}} ~.
\end{align}
For the K model ($\eta = 1$, $\beta = 1$)
\begin{align}
\label{eq:kinoverdenselimit}
    \sin^2(2\theta)_\textrm{DW,lim}^{\rm K} ~=&~ 7.03\times10^{-6} \left(\frac{m_s}{\textrm{keV}}\right)^{-\frac{5}{3}}\left(\frac{T_{\rm tr}}{5 \textrm{ MeV}}\right)^{-1} \left(\frac{T_{\nu, 0}}{1.95~{\rm K}}\right)^{-3}\left(\frac{g_{\ast}}{30}\right)^{\frac{3}{2}} \notag\\ &\times\left(\frac{d_\alpha}{1.13}\right)^{-1} \left(\frac{\Omega_\textrm{DM}h^2}{ 0.12}\right)^{-1}\left(\frac{B}{10.88\times10^{-9}\textrm{ GeV}^{-4}}\right)^{\frac{1}{3}} ~.
\end{align}
For ST2 ($\eta = 0.03$, $\beta = 0$) instead,
\begin{align}
\label{eq:st2overdenselimit}
    \sin^2(2\theta)_\textrm{DW,lim}^{\rm ST2} ~=&~ 1.29\times10^{-8} \left(\frac{m_s}{\textrm{keV}}\right)^{-2} \left(\frac{T_{\nu, 0}}{1.95~{\rm K}}\right)^{-3} \left(\frac{g_{\ast}}{30}\right)^{\frac{3}{2}}\left(\frac{d_\alpha}{1.13}\right)^{-1}\notag\\ &\times\left(\frac{\Omega_\textrm{DM}h^2}{ 0.12}\right)^{-1} \left(\frac{B}{10.88\times10^{-9}\textrm{ GeV}^{-4}}\right)^{\frac{1}{2}} ~.
\end{align}
For the LRT model, with $T_{\rm tr} = T_{\rm RH}$,
\begin{align}
\label{eq:lrtoverdenselimit}
    \sin^2(2\theta)_\textrm{DW,lim}^{\rm LRT} ~=~ 1\times10^{-3}\left(\frac{m_s}{\textrm{keV}}\right)^{-1}\left(\frac{T_{\rm tr}}{5 \textrm{ MeV}}\right)^{-3} \left(\frac{T_{\nu, 0}}{1.95~{\rm K}}\right)^{-3}\left(\frac{d_\alpha}{1.13}\right)^{-1}\left(\frac{\Omega_\textrm{DM}h^2}{0.12}\right)^{-1} ~.
\end{align}

\clearpage
\bibliography{sternumodcos}
\addcontentsline{toc}{section}{Bibliography}
\bibliographystyle{JHEP}
\end{document}